\documentclass[11pt]{article}

\renewcommand{\arraystretch}{1.3}
\makeatletter
\newdimen\normalarrayskip              
\newdimen\minarrayskip                 
\normalarrayskip\baselineskip
\minarrayskip\jot
\newif\ifold             \oldtrue            \def\new{\oldfalse}
\def\arraymode{\ifold\relax\else\displaystyle\fi} 
\def\eqnumphantom{\phantom{(\theequation)}}     
\def\@arrayskip{\ifold\baselineskip\z@\lineskip\z@
     \else
     \baselineskip\minarrayskip\lineskip2\minarrayskip\fi}
\def\@arrayclassz{\ifcase \@lastchclass \@acolampacol \or
\@ampacol \or \or \or \@addamp \or
   \@acolampacol \or \@firstampfalse \@acol \fi
\edef\@preamble{\@preamble
  \ifcase \@chnum
     \hfil$\relax\arraymode\@sharp$\hfil
     \or $\relax\arraymode\@sharp$\hfil
     \or \hfil$\relax\arraymode\@sharp$\fi}}
\def\@array[#1]#2{\setbox\@arstrutbox=\hbox{\vrule
     height\arraystretch \ht\strutbox
     depth\arraystretch \dp\strutbox
     width\z@}\@mkpream{#2}\edef\@preamble{\halign
\noexpand\@halignto
\bgroup \tabskip\z@ \@arstrut \@preamble \tabskip\z@ \cr}%
\let\@startpbox\@@startpbox \let\@endpbox\@@endpbox
  \if #1t\vtop \else \if#1b\vbox \else \vcenter \fi\fi
  \bgroup \let\par\relax
  \let\@sharp##\let\protect\relax
  \@arrayskip\@preamble}
\def\eqnarray{\stepcounter{equation}%
              \let\@currentlabel=\theequation
              \global\@eqnswtrue
              \global\@eqcnt\z@
              \tabskip\@centering
              \let\\=\@eqncr
 \halign to \displaywidth\bgroup
    \eqnumphantom\@eqnsel\hskip\@centering
    $\displaystyle \tabskip\z@ {##}$%
    \global\@eqcnt\@ne \hskip 2\arraycolsep
         $\displaystyle\arraymode{##}$\hfil
    \global\@eqcnt\tw@ \hskip 2\arraycolsep
         $\displaystyle\tabskip\z@{##}$\hfil
         \tabskip\@centering
    &{##}\tabskip\z@\cr}
\begingroup\ifx\undefined\newsymbol \else\def\input#1 {\endgroup}\fi

\newcounter{app}

\def\app{\setcounter{equation}{0}
\def\theequation{A\Roman{app}.\arabic{equation}}\par
   \addvspace{4ex}
   \@afterindentfalse
  \secdef\@app\@dapp}
\newcommand\@app{\@startsection {app}{1}{0ex}%
                                   {-3.5ex \@plus -1ex \@minus -.2ex}%
                                   {2.3ex \@plus.2ex}%
                                   {\normalfont\Large\bf}}

\def\@dapp#1{%
{\parindent \z@ \raggedright  \bf #1}\par\nobreak}
\def\l@app#1#2{\ifnum \c@tocdepth >\z@
    \addpenalty\@secpenalty
    \addvspace{1.0em \@plus\p@}%
    \setlength\@tempdima{8.5em}%
    \begingroup
      \parindent \z@ \rightskip \@pnumwidth
      \parfillskip -\@pnumwidth
      \leavevmode \bfseries
      \advance\leftskip\@tempdima
      \hskip -\leftskip
      #1\nobreak\hfil \nobreak\hb@xt@\@pnumwidth{\hss #2}\par
    \endgroup\fi}
\newcounter{sapp}[app]

\def\sapp{\def\theequation{A\arabic{app}.\arabic{equation}}\par
   \@afterindentfalse
  \secdef\@sapp\@dsapp}
\newcommand\@sapp{\@startsection{sapp}{2}{\z@}%
                                     {-3.25ex\@plus -1ex \@minus -.2ex}%
                                     {1.5ex \@plus .2ex}%
                                     {\normalfont\large\bfseries}}

\def\@dsapp#1{%
{\parindent \z@ \raggedright  \bf #1}\par\nobreak}
\newcommand{\l@sapp}{\@dottedtocline{2}{1.5em}{3em}}

\catcode`\@=11
\def\marginnote#1{}

\newcount\hour
\newcount\minute
\newtoks\amorpm
\hour=\time\divide\hour by60
\minute=\time{\multiply\hour by60 \global\advance\minute by-\hour}
\edef\standardtime{{\ifnum\hour<12 \global\amorpm={am}%
        \else\global\amorpm={pm}\advance\hour by-12 \fi
        \ifnum\hour=0 \hour=12 \fi
        \number\hour:\ifnum\minute<10 0\fi\number\minute\the\amorpm}}
\edef\militarytime{\number\hour:\ifnum\minute<10 0\fi\number\minute}

\def\draftlabel#1{{\@bsphack\if@filesw {\let\thepage\relax
      \xdef\@gtempa{\write\@auxout{\string
          \newlabel{#1}{{\@currentlabel}{\thepage}}}}}\@gtempa \if@nobreak
    \ifvmode\nobreak\fi\fi\fi\@esphack} \gdef\@eqnlabel{#1}}
    \def\@eqnlabel{}
\def\@vacuum{}
\def\draftmarginnote#1{\marginpar{\raggedright\scriptsize\tt#1}}

\def\draft{
  \oddsidemargin -.5truein
  \def\@oddfoot{\footnotesize \sl preliminary draft \hfil
    \rm\thepage\hfil\sl\today\quad\militarytime}
  \let\@evenfoot\@oddfoot \overfullrule 3pt
    \let\label=\draftlabel
    \let\marginnote=\draftmarginnote
  \def\@eqnnum{(\theequation)\rlap{\kern\marginparsep\tt\@eqnlabel}%
    \global\let\@eqnlabel\@vacuum}

  }

\voffset= - 1.2in
\hoffset= - 1.0in
\def\be{\begin{eqnarray}}
\def\ee{\end{eqnarray}}
\def\nn{\nonumber}

\def\p{\partial}

\def\beq{\begin{equation}}
\def\eeq{\end{equation}}
\def\ba{\beq\new\begin{array}{c}}
\def\ea{\end{array}\eeq}
\def\be{\ba}
\def\ee{\ea}

\def\theequation{\arabic{section}.\arabic{equation}}

\newfont{\alef}{msbm10 at 11pt}
\newfont {\goth}{eufm10 at 11pt}
\def\mathbb#1{\hbox{{\alef #1}}}

\def\tr{\hbox{tr }}
\def\Tr{\hbox{Tr }}

\let\@@savethanks\thanks
\def\thanks#1{\gdef\thefootnote{\alph{footnote}}\@@savethanks{#1}}

\title{{\bf BGWM  as Second Constituent of Complex Matrix
Model
}
\vspace{.5cm}}
\author{{\bf A.Alexandrov}\thanks{E-mail: \ al@itep.ru}
\date{ } \\ {\small
{\it Blackett
Laboratory, Imperial College, London SW7 2AZ, U.K.}
and {\it ITEP, Moscow, Russia}}\\ \\
{\bf A. Mironov}\footnote{E-mail:
\ mironov@itep.ru; mironov@lpi.ru}
\date{ } \\
{\small {\it Lebedev Physics Institute}
and {\it ITEP, Moscow, Russia}}\\ \\
{\bf A.Morozov}\thanks{E-mail: \ morozov@itep.ru}
\date{ } \\ {\small {\it ITEP, Moscow, Russia}}
}

\begin{document}

\maketitle

\vspace{-10.5cm}

\begin{center}
\hfill FIAN/TD-8/09\\
\hfill Imperial-TP-AA-2009-1\\
\hfill ITEP/TH-25/09\\
\end{center}

\vspace{9.0cm}

\begin{abstract}
\noindent In \cite{AMMIM} we explained that partition functions
of various matrix models can be constructed from that of the cubic
Kontsevich model, which, therefore, becomes a basic elementary
building block in "M-theory" of matrix models \cite{AMMMT}. However,
the less topical complex matrix model appeared to be an exception:
its decomposition involved not only the Kontsevich $\tau$-function
but also another constituent, which we now identify as the
Brezin-Gross-Witten (BGW) partition function. The BGW
$\tau$-function can be represented either as a generating function
of all unitary-matrix integrals or as a Kontsevich-Penner model with
potential $1/X$ (instead of $X^3$ in the cubic Kontsevich model).
\end{abstract}
\newpage

\tableofcontents

\newpage
\def\thefootnote{\arabic{footnote}}

\section*{Introduction}
\def\theequation{\arabic{equation}}
\setcounter{equation}{0}

Matrix models \cite{hemamo,MaMo} play a very special role
in modern theoretical physics.
They appear regularly and prove useful
in analysis of various simplified models
of concrete physical phenomena,
but their real significance
is that they somehow capture and reflect the very basic properties of
string theory -- and can serve to represent the universal classes
of quantum field theory models.
From the very beginning matrix models were introduced to describe some
very general features (eigenvalue repulsion) of statistical
distributions \cite{Dy}.
It was, perhaps, the first recognition of the role of group theory
-- the underlying theory behind matrix models --
in explaining the fundamental properties of quantum/statistical behavior.
Much later this led to discovery that integrability is the basic
property of all functional integrals, considered as functionals on
the moduli space of theories \cite{UFN2}, and matrix models played
a central role \cite{UFN3} in the formulation of the fundamental relation
\be
{\framebox{partition function = $\tau$-function}}
\label{pftf}
\ee
between the two central concepts of modern theory, already with a variety
of applications in different fields, from gauge theories \cite{SW}
to Hurwitz theory \cite{HT} and with still
many more to come.
An immediate implication of (\ref{pftf}) is that quantitative approach
-- a possibility to {\it calculate} something --
in string theory (= a theory of {\it families} of quantum mechanical models)
requires extension of the standard set of special functions to
a broader set of $\tau$-functions \cite{tauf,gentau} --
a far-going generalization
of both hypergeometric and elliptic families.
A highly non-trivial step here was introduction of "infinite-genus"
$\tau$-functions, satisfying the string equations \cite{Witogra} and
Virasoro/$W$-constraints \cite{FKN}-\cite{ShaW},
and it was once again inspired by the study of matrix models.
Unfortunately, even the simplest of these $\tau$-functions, associated
with Hermitian \cite{hemamo} and Kontsevich \cite{Ko,GKM,versus,komo} matrix models,
are not yet systematically studied/tabulated and still can not be included
into the special-functions textbooks -- see \cite{spef} for the first
attempts in this direction.
It is very important to realize that the world of such $\tau$-functions
is cognizable, and is, perhaps, actually finitely-generated:
many (all?) matrix-model $\tau$-functions are actually expressed by
group-theoretical methods through a few basic ones.
This decomposition results from description of genus expansion of
matrix-model partition functions in terms of auxiliary "spectral"
Riemann surfaces \cite{Giv,spef,Ey}, and explicitly relates them to
representation theory of Krichever-Novikov type deformation \cite{KN}
of Kac-Moody algebras.
One of spectacular byproducts of this development is the possibility
to build a "string-field-theory-like" diagram technique \cite{Ey}
for the model of entire string theory, provided by
"M-theory of matrix models" \cite{AMMMT}.
As shown in \cite{AMMIM}, the main basic block (constituent) of the
matrix model partition functions in this approach is the ordinary
Kontsevich $\tau$-function $Z_K$ of \cite{Ko}.

However, already in \cite{AMMIM} a first counter-example was found to
this (over?)-optimistic conjecture: partition function of the
complex matrix model \cite{comamo,compen} is made not only from $Z_K$,
but also from some other ingredient, denoted $\tilde Z_K$ in
s.8 of \cite{AMMIM}.
The purpose of the present paper is to identify this $\tilde Z_K$
with a very important and well-known partition function: that of
Brezin-Gross-Witten model (BGWM) \cite{BGW,GN,GKMU}:
\be
\framebox{$\tilde Z_K = Z_{BGW}$}
\ee
By definition, BGWM describes correlators of unitary matrices
with a non-linear Haar measure.
Unitary correlators play a crucially important role in description
of gluons in lattice gauge-theory models \cite{latthe}, however,
unitary matrix models are more complicated than Hermitian ones, they
are in intermediate position between eigenvalue and non-eigenvalue
models and remain under-investigated, see \cite{unif}-\cite{unil} for
some crucial references.
A modern matrix-model-theory approach to BGWM and its embedding into
the set of  generalized Kontsevich models (GKM) \cite{GKM}
was outlined in \cite{GKMU},
but has not been developed any further since then.
Hopefully reappearance of this model in the context of matrix-model
M-theory will help to attract new attention to this
unjustly-abandoned subject.

\bigskip

We begin in s.\ref{4mamo} from reminding the definition and
the main properties of the four partition functions which
participate in our story:
$Z_H(t)$, $Z_C(t)$, $Z_K(\tau)$ and $Z_{BGW}(\tau)$.
They were originally introduced
as matrix integrals over
Hermitian ($Z_H$ and $Z_K$), complex ($Z_C$) and unitary ($Z_{BGW}$)
matrices, with the time-variables identified either with the coupling
constants ($t_k$ in $Z_H$ and $Z_C$) or with the Miwa transform
of the background field ($\tau_k$ in $Z_K$ and $Z_{BGW}$).

As functions of their parameters -- the time-variables $t$ or $\tau$ --
these integrals satisfy Ward identities (or Picard-Fucks equations)
\cite{vircMM}, which have the form of the Virasoro constraints.
Namely,
\be
\frac{\partial Z_H}{\partial t_k} = \hat L_{k-2} Z_H, \ \ \ \ \
\frac{\partial Z_C}{\partial t_k} = \hat L_{k-1} Z_C
\label{dico}
\ee
with $k\geq 1$,  with $\frac{\partial Z_H}{\partial t_0} = N/g Z_H$,
$\frac{\partial Z_C}{\partial t_0} = N/g Z_C$, where $N$ is the size
of the matrix in the original integral representation
and with "discrete-Virasoro"
operators \cite{GMMMO}
\be
\hat L_m = \sum_{k=1}^\infty kt_k\frac{\partial}{\partial t_{m+k}}
+ g^2\!\!\!\!\sum_{\stackrel{a,b=0}{a+b=m}}^m
\frac{\partial^2}{\partial t_a\partial t_b},\ \ \ \ m\geq -1
\label{divi}
\ee
and
\be
\frac{\partial Z_K}{\partial \tau_{k-1}} = \hat {\cal L}_{k-2} Z_K, \ \ \ \ \
\frac{\partial Z_{BGW}}{\partial \tau_{k-1}} = \hat {\cal L}_{k-1} Z_{BGW}
\label{coco}
\ee
with $k\geq 1$ and with "continuous-Virasoro" operators \cite{FKN,DVV}
\be
\hat {\cal L}_m = \sum_{k=1}^\infty \left(k+\frac{1}{2}\right)
\tau_k\frac{\partial}{\partial \tau_{m+k}}
+ g^2\!\!\!\!\sum_{\stackrel{a,b=0}{a+b=m-1}}^{m-1}
\frac{\partial^2}{\partial \tau_a\partial \tau_b}
+ \frac{\tau_0^2}{16g^2}\delta_{m,-1} + \frac{1}{16}\delta_{m,0},
\ \ \ \ m\geq -1
\label{covi}
\ee
Now one can switch to $D$-module approach and define the
four partition functions as solutions to the four systems
of linear differential equations (\ref{dico}) and (\ref{coco}),
and original integral formulas
are just integral representations for the solutions.

Not surprisingly,
such representations are not unique, and one can instead
represent the same solutions in a very different integral form:
of Kontsevich-Penner integrals over $n\times n$ Hermitian matrices
with a peculiar Penner term $N\tr\! \log \phi$ in the action.
This puts all the four models in the unifying context of GKM theory
\cite{GKM}.
Direct relation between the two integral
representations
is provided by a version of Faddeev-Popov trick from \cite{ShaW}.

All four matrix integrals can be expressed in the form of determinants
of the other matrices, which have ordinary single integrals
as their elements. These determinant representations are very
important, because they are typical for the {\it tau-functions}
of integrable hierarchies \cite{KMMOZ,versus}
-- the generalized characters of Lie algebras \cite{GKMKM,gentau}.
In other words, partition functions of the matrix models are
always the tau-functions \cite{UFN3}. Moreover, this is a general
property of all partition functions -- the generating functions of
all correlation functions in any quantum theory, -- this is a
consequence of the freedom to change integration variables (fields)
in the functional integral \cite{UFN2}.
For tau-functions the whole sets of Virasoro constraints
are actually fixed by their lowest components $\hat L_{-1}$
or $\hat L_0$, which therefore has its own name: the
{\it string equation} \cite{Witogra}.

Integrability means that partition function satisfies a bilinear
Hirota equation \cite{tauf} of the form \be
\oint Z_N\!\left(t_k + \frac{1}{kz^k}\right) Z_{N'}\!\left(t'_k -
\frac{1}{kz^k}\right)z^{N'-N} e^{\sum_k (t'_k-t_k)z^k} \ dz \ee
This equation has its origin in decomposition rule $R\times R' =
\sum_I R_I$ for representations of Lie algebras and this is an
equation for the {\it characters} of the algebra
\cite{GKMKM,gentau}. For loop algebras the characters can be rather
non-trivial, they can be actually labeled by some auxiliary
(spectral) Riemann surfaces or, better, by the points of an
infinite-dimensional Grassmannian (the {\it universal moduli space}
of \cite{UMS}) -- what means that the spectral surface can actually
have an infinite genus (and this is typically the case for the
matrix-model partition functions).

For irreducible representations of finite-dimensional simple Lie
algebras, the characters are given by two determinant Weyl
formulas: in the case of $SL(N)$ and representation labeled by
partition $\vec m:\ m_1\geq m_2\geq\ldots\geq m_N\geq0$ (the
Young diagram) \be \chi_{\vec m}(t) = \det P_{i+m_j-1}(t),
\label{fWf} \ee where $P_i(t)$ are the Shur polynomials,
$\exp\left(\sum_{k} t_kx^k\right) = \sum_m x^mP_m(t)$, and, after
the Miwa transform $t_k = \frac{1}{k}\sum_i \lambda_i^{-k}$, \be
\chi_{\vec m}(t) = \frac{\det_{ij} \lambda_i^{j +
m_j}}{\Delta(\lambda)}, \label{sWf} \ee

In fact, the two Weyl formulas are mirrored in two possible
representations of matrix models partition function: as
we shall demonstrate, the Hermitian and complex matrix models,
besides the standard determinant representation of type (\ref{fWf}), have
a Kontsevich-Penner representation of type (\ref{sWf}).

Virasoro constraints can be used as recursion relations to
provide the logarithm of the partition functions $g^2\log Z$ in the form of the formal series
in non-negative powers of $t$-variables and $g^2$.
Such formal series are unambiguously defined
by the systems (\ref{dico}) and (\ref{coco}), since $k\geq 1$
in all the four cases.
The generating functions ("multiresolvents") $\rho^{(\!p|q)}(z)$,
defined as
\be
\rho^{(\cdot|q)}(z_1,\ldots,z_q) = \hat\nabla(z_1)\ldots\hat\nabla(z_q)\log Z\big|_{t=0}
=\  \sum_{p=0}^\infty g^{2p-2}\rho^{(\!p|q)}(z_1,\ldots,z_q)
\label{muddef}
\ee
possess an important property: they are poly-differentials on
auxiliary {\it spectral} Riemann surfaces (complex curves)
\cite{spef,Ey}, which for the four matrix models in question are
all double-coverings of the Riemann sphere with only two ramification points (and thus Riemann
spheres themselves).
The spectral curve representation arises only for the special
choice of generating functions: they should be {\textit{resolvents}},
i.e. no $k$-dependent coefficients are allowed in (\ref{muddef}).
There are other interesting choices of coefficients, when
alternative generating functions possess other interesting
properties, see for example \cite{HZ}.

\bigskip

In s.\ref{decof} we proceed to decomposition formulas.
Virasoro constraints can be also considered as quadratic differentials
on the spectral surfaces, expanded near particular points.
It turns out that "discrete" operators (\ref{divi}) arise in
expansion near non-singular points, while "continuous"
operators (\ref{covi}) -- in those near ramification points
of degree $2$. This means that a globally-defined Virasoro
quadratic differential can be decomposed in both bases
and this idea finally leads to decomposition formulas \cite{AMMIM}.

The basic one is
\be
Z_H = \hat U_{KK} (Z_K\otimes Z_K),
\label{dehekk}
\ee
it expresses $Z_H(t)$ for Hermitian matrix model through
$Z_C(\tau)$ for Kontsevich model. It is explained in full detail
in s.\ref{H}. Ingredients of the construction are:
explicit parametrization of the spectral curve and of the
singular differentials in the vicinities of particular points
on it, explicit formula for the global $\hat U(1)$ current
and the Virasoro differential, its projection onto "canonical"
quadratic differentials in the vicinities of the particular points
and  Bogoliubov transform of the time-variables with the help of
the conjugation $\hat U$-operator, and, finally,
projection from generic Laurent
series for the Virasoro operator to a Taylor series, provided
by peculiar projection operator ${\cal P}$, which picks up
a triangular subalgebra from entire Virasoro (Krichever-Novikov)
algebra. Generators of this triangular subalgebra can be imposed as constraints
on partition function and form a consistent and resolvable set
of constraints.

From the point of view of $D$-module approach
the difference between $Z_H$ and $Z_C$ in (\ref{dico}) looks minor:
both are annihilated
by {\it the same} discrete-Virasoro operators $\hat L_n$,
only $n\geq -1$ for $Z_H$, but $n\geq 0$ for $Z_C$.
The second difference is that the shift of
time-variables, which generates the l.h.s. in (\ref{dico}),
is also different: $t_2$ is shifted in the case
of $Z_H$, but $t_1$ is shifted in the case of $Z_C$ -- this is
important to explain why both sets of equations (\ref{dico}) have
unambiguous formal-series solution,
despite the set of constraints contains one less equation in the
case of $Z_C$. However, this second difference is not essential
for comparison of (\ref{dico}) and (\ref{coco}). Indeed, the
relation between $Z_K$ and $\tilde Z_K = Z_{BGW}$ in (\ref{coco})
is exactly the same:
both partition functions are annihilated by {\it the same}
continuous-Virasoro operators ${\cal L}_n$, but $n\geq -1$ for
$Z_K$, while $n\geq 0$ for $Z_{BGW}$.
This time shifted are $\tau_1$ in $Z_K$ and $\tau_0$ in
$Z_{BGW}$, what guarantees that the formal-series solutions
are unambiguously defined in both cases.
All this implies that in the spectral-surface formalism
the difference between $Z_H$ and $Z_C$ is concentrated in the
choice of projection operator ${\cal P}$ at the last stage.
Since projection operator can be realized as a contour integral,
its modification can actually be shifted from $Z_H$
(where it transformed $Z_H$ into $Z_C$) to one of the two $Z_K$
at the r.h.s. of (\ref{dehekk}) and convert it into
$\tilde Z_K = Z_{BGW}$. In other words, instead of (\ref{dehekk})
we obtain
\be
Z_C = \hat U_{K\tilde K} (Z_K\otimes\tilde Z_K)
\ \ \ \ {\rm i.e.} \ \ \ \
Z_C = \hat U_{K\tilde K} (Z_K\otimes Z_{BGW})
\label{decoc}
\ee
This is the main result of the present paper and it is
discussed in full detail in s.\ref{CKBGW}.
This formula was supported in \cite{AMMIM}
by explicit comparison of the first
terms of expansions of partition functions at both sides of
(\ref{decoc}).

\bigskip

Some concluding remarks are given in s.\ref{concl}.

\pagebreak

\part{The four matrix models \label{4mamo}}

\setcounter{equation}{0}
\def\theequation{\Roman{part}.\arabic{equation}}

\section{Hermitian matrix model $Z_H(t)$}

\subsection{Original integral representation}

Partition function of Hermitian matrix model \cite{hemamo} is
defined by the integral over hermitian $N\times N$ matrix
\be
Z_H(t) = \int_{N\times N} \exp
\left(-\frac{1}{2g}\Tr H^2
+ \frac{1}{g}\!\sum_{k=0}^\infty  t_k\Tr H^k \right) dH
\label{Horin}
\ee
where the measure $dH = \prod_{i,j=1}^N dH_{ij}$.
This is nothing but the generating function of all
$GL(N)$-invariant Gaussian correlators of Hermitian matrix $H$.

\subsection{Eigenvalue representation}

The Hermitian matrix $\Phi$ can be diagonalized by a unitary
transformation, $H = UDU^\dagger$, where $D = {\rm diag}(H_1,
\ldots, H_N)$ is the diagonal matrix made from the eigenvalues of $H$.
The norm of $H$ decomposes as \be \label{Hmeas0}\Tr (\delta H)^2 = \Tr (\delta
D)^2 + \Tr \left(\left[ U^\dagger \delta U, D\right]\right)^2 =
\sum_{i=1}^N (\delta H_i)^2 +
\sum_{i<j}^N (H_i-H_j)^2
(U^\dagger\delta U)_{ij}^2
\ee so that the measure \be dH = [dU] \prod_{i<j}^N(H_i-H_j)^2
\prod_{i=1}^N dH_i \label{Hmeas} \ee where $[dU]$ is the non-linear
(Haar)
measure for unitary matrices. Since the action in (\ref{Horin}) does
not depend on $U$, the integral $V_N = \int [dU]$ fully decouples, and (\ref{Horin})
turns into an $N$-fold integral over eigenvalues $H_i$ with the peculiar
square of the Van-der-Monde determinant $\Delta(H) =
\prod_{i>j}^N(H_i-H_j)$ in the measure:
\be\label{eiint}
Z_H(t)=V_N\int\prod_idH_i\Delta^2 (H)\exp\left(
-{1\over 2g}\sum_iH_i^2+{1\over g}\sum_{i,k}t_kH_i^k\right)
\ee

\subsection{Virasoro constraints}

Integral (\ref{Horin}) is invariant under any change
of integration matrix-variable $H$. In particular, $Z_H(t)$ does
not change if one substitutes $H \rightarrow H + \epsilon H^{n+1}$
with any matrix-valued parameter $\epsilon$ and any integer
$n\geq -1$. This invariance implies that \cite{vircMM}
\be
\frac{\partial Z_H(t)}{\partial t_{n+2}} = \hat L_n Z_H(t),
\ \ \ \ n\geq -1
\label{Hdico}
\ee
with the operator $\hat L_n$ defined in (\ref{divi}) and
\be
\frac{\partial Z_H(t)}{\partial t_0} = \frac{N}{g}Z_H(t)
\label{Ht0}
\ee
The l.h.s. in (\ref{Hdico}) is produced by the shift of
the $t_2$ variable $t_2 \rightarrow t_2 - 1/2$ in the
initial formula (\ref{Horin}).

Together with (\ref{Ht0})
the system (\ref{Hdico}) provides a set of recurrent relations
which allows one to unambiguously construct $Z_H(t)$ term-by-term as
a formal series in non-negative powers of $t$-variables.

\subsection{Determinant representations and integrability\label{Hdere}}

As we shall explain now, the properly normalized matrix integral
$\displaystyle{{\cal Z}_N\equiv {1\over V_NN!}Z_H(t)}$
is a $\tau$-function of the Toda-chain integrable hierarchy
\cite{GMMMO,UFN3}. In this paragraph we rescale the time variables to cancel
the coefficient $g$ in front of them in order to have the standard definition of
integrable hierarchies.

One of the technical ways to deal with integrals of form
(\ref{eiint}) was proposed in \cite{hemamo}. The authors introduced a
system of orthogonal polynomials with the orthogonality condition
\be \label{OP} \int P_i(H)P_j(H)e^{-V(H)}dH = \delta _{ij}e^{\varphi
_i(t)}, \ee
where $e^{\varphi _i(t)}$ are norms defined by integral
(\ref{OP}) and the normalizing condition for the polynomials is
\be
\label{OPN} P_i(H) = \sum _{j\leq i}\gamma_{ij}H^j , \ \ \gamma_{ii}
= 1, \ee
i.e., the coefficient of the leading term is put equal to
unity.

Using polynomials (\ref{OP}), (\ref{OPN}), we can rewrite
(\ref{eiint}) as
\be \label{prod} {\cal Z}_N =  (N!)^{-1}\int
\prod  _i dH_i \det P_{k-1}(H_j) \det P_{l-1}(H_m) \exp \{ -{1\over 2}\sum_iH^2_i+\sum
_{i,k}t_{_k}H^{_k}_i\} = \prod ^{N-1}_{i=0}e^{\varphi _i(t)}.
\ee

In order to get the determinant representation, we rewrite
orthogonality condition (\ref{OP}) in a ``matrix" form. That is, we
introduce the matrix $\Gamma$ with matrix elements $\gamma_{mn}$
defined by (\ref{OPN}), the so-called {\it moment matrix} $C$ with
the matrix elements
\be \label{MomMat} C_{ij}= \int dH H^{i+j-2}
e^{-V(H)},
\ee
and the diagonal matrix $J$ with diagonal elements $
e^{\varphi _n }$. Then, (\ref{OP}) can be written as a matrix
relation\footnote{This relation is nothing but the Riemann--Hilbert
problem also known as the factorization problem, see
\cite{UT,GMMMO,SemT} for the details.}
\be \Gamma C\Gamma^T=J
\ee
where $\Gamma^T$ is the transposed matrix. Evaluating the
determinant of the both sides of this relation and using
(\ref{prod}), we obtain
\be \label{detrepTC} {\cal Z}_N =
\det_{N\times N} C_{ij}.
\ee
The moment matrix satisfies a number
of relations, which follow directly from its explicit form
(\ref{MomMat}),
\be\label{cond1} {\partial C_{\ast}(t)\over
\partial t_k} \equiv \partial_k  C_{\ast}(t) =  {\partial^k
C_{\ast}(t)\over \partial t_1^k} \equiv \partial^k
C_{\ast}(t), \ee
\be \label{cond2} C_{ij} = C_{i+j}
\ee
and
\be
\label{cond3} C_N = \partial^{N-2} C_{11} \equiv
\partial^{N-2} C
\ee
Finally, the partition function of the
one-matrix model is
\be \label{detrepTC2} {\cal Z}_N = \det
\partial ^{i+j-2} C,
\ee
which results in the Toda chain
\cite{versus,KMMOZ,UFN3,GMMMO}. Note that conditions (\ref{cond1})
and (\ref{cond3}) are satisfied for the whole hierarchy of the
two-dimensional Toda lattice and for the KP hierarchy, while
(\ref{cond2}) is specific for the Toda chain.

\subsection{Kontsevich-Penner representation \label{Hpen}}

The system (\ref{Hdico}) is solved by another integral, very
different from (\ref{Horin}) \cite{CheGK}: \be Z_H(t) =
\pi^{\frac{N^2-n^2}{2}}g^{\frac{(N-n)^2}{2}}i^{-Nn}e^{\frac{1}{2g}\tr L^2} \int_{n\times n}
\exp \left( -\frac{g}{2}\tr h^2
 +N\tr\log h-i\tr h L\right)
dh
\ \ \ \ \
\label{PKH}
\ee
Integral is now over $n\times n$ Hermitian matrix $h$,
$dh = \prod_{a,b=1}^n dh_{ab}$ and depends on
additional $n\times n$ matrix (background field) $L$.
To emphasize the difference between $N$ and $n$ we use
small letters for $h$ and ${\rm tr}$ instead of
$H$ and ${\rm Tr}$ in (\ref{Horin}).
If expanded around a saddle-point $h=L$ this integral
is a formal series in positive powers of variables
\be
t_k = -\frac{g}{k}\tr L^{-k}, \ \ \ k\geq 1
\label{HMit}
\ee
and\footnote{It is possible to introduce dependence on $t_0$ implicitly, namely to define the partition function as follows:
\be
Z_H(t) =
\pi^{\frac{N^2-n^2}{2}}g^{\frac{(N-n)^2}{2}}i^{-Nn}e^{\frac{t_0N}{g}+\frac{1}{2g}\tr L^2-N tr \log L}
 \int_{n\times n}
\exp \left( -\frac{g}{2}\tr h^2
 +N\tr\log h-i\tr h L\right)dh
\ee
this slightly change all calculations but leave the partition function unchanged.}
\be\label{HMit0}
t_0=g \ \tr \log L
\ee
This is a model from the GKM family and peculiar logarithmic
term in the action is often named Penner term \cite{Pen},
so that (\ref{PKH}) is known as
{\it Gaussian Kontsevich-Penner model}.

\subsubsection{Proof I: Ward identities}

In order to check that (\ref{PKH}) satisfies (\ref{dico})
one begins with the Ward identity for this integral \cite{GKM},
associated to the shift $h \rightarrow h + \epsilon$
of the integration variable by a small arbitrary matrix $\epsilon$:
\be
\left(g\frac{\p}{\p L_{tr}}+N\left(\frac{\p}{\p L_{tr}}\right)^{-1}+L\right)
\int_{n\times n} \exp \left( -\frac{g}{2}\tr h^2
 +N\tr\log h-i\tr h L\right)=0
\label{HGN} \ee which gives \be \left(g\left(\frac{\p}{\p
L_{tr}}-\frac{L}{g}\right)^2+N+n+L\left(\frac{\p}{\p
L_{tr}}-\frac{L}{g}\right)\right)Z=\left(g\frac{\p^2}{\p
L_{tr}^2}+N-L\frac{\p}{\p
L_{tr}}\right)Z=0 \ee
It remains to substitute a
function $Z_H(t)$ with $t$'s expressed through $L$ by the {\it Miwa
transform} (\ref{HMit}). Then (\ref{HGN}) becomes (\ref{Hdico}), see
\cite{CheGK, versus,UFN3} for technical details.

\subsubsection{Proof II: Orthogonal polynomials\label{orthopol}}

The two integral representations (\ref{Horin}) and
(\ref{PKH}) can be related directly, without a reference
to Virasoro constraints (\ref{Hdico}).
One procedure, making use of orthogonal polynomials
(Hermite polynomials in this particular case) is described
in details in \cite{versus,UFN3}. One can rewrite (\ref{prod}) with time variables
substituted by (\ref{HMit}), (\ref{HMit0}) in terms of orthogonal polynomials
in the following way:
\be
{\cal Z}_N=(N!)^{-1}\int   \prod  _i dH_i \Delta ^2(H) \exp
\left(-\sum_i\tilde V(H_i)+{1\over g}\sum _{i,k=0}t_{_k}H^{_k}_i \right)=\nn\\
= (N!)^{-1}\int   \prod  _i dH_i \Delta ^2(H)\exp\left(-\sum_i\tilde V(H_i)\right)
\prod _{i,a}(L_a -
H_i) = \nn\\
=(N!)^{-1}\int   \prod  _i dH_i \exp\left(-\sum_i\tilde V(H_i)\right)
\Delta (H)
{\Delta (H,L)\over \Delta (L)}=\nn\\
= (N!)^{-1}\Delta ^{-1}(L)\int   \prod  _i dH_i
\exp\left(-\sum_i\tilde V(H_i)\right) \times \nn\\
\times  \det _{N\times N} \tilde P_{i-1}(H_j) \det _{(N+n)\times (N+n)}
\left[ \begin{array}{rcl}
\tilde P_{i-1}(H_j) & \vdots & \tilde P_{N+b-l} (H_j)\\
\ldots & \vdots & \ldots\\
\tilde P_{i-1}(L_a) & \vdots & \tilde P_{N+b-l} (L_a)
\end{array}
\right]\hbox{ , }
\ee
where $i,j=1,...,N$, $a,b=1,...,n$ and the orthogonal polynomials $\tilde P_{k}(H)$
are orthogonal with the measure $\exp(-\tilde V(H))$. In the case under consideration,
it is equal to $\exp\left(-{1\over 2g}H^2\right)$. Now calculating the determinants and using
the orthogonality condition (\ref{OP}), one arrives at
\be\label{DMM}
{\cal Z}_N = \Delta ^{-1}(L) \det _{n\times n}
\tilde P_{N+a-1}(L_b) \prod  _i e^{\tilde \varphi _i(s)} = \nn\\
= \left[ \prod  _i e^{\tilde \varphi _i(s)}\right]
{\det _{(ab)}\phi ^{(N)}_a(L_b)\over \Delta (L)} = \left.{\cal Z}_N\right|_{L_a=\infty}
\times {\det _{(ab)}\phi ^{(N)}_a(L_b)\over \Delta (L)}\hbox{ ,}
\ee
with
\be\label{v3.24}
\phi ^{(N)}_a(L) = \tilde P_{N+a-1}(L)
\ee

Now let us see that integral (\ref{PKH}) can be also transformed to this form. To this end,
we need the Itzykson-Zuber formula, \cite{IZ}
\be
\int_{U(n)}dUe^{\tr AUBU^\dag}=V_n{\det e^{a_ib_j}\over\Delta (x)\Delta (y)}
\ee
where integral runs over unitary $n\times n$ matrices with the Haar measure $dU$, and $a_i$,
$b_j$ are eigenvalues of Hermitian matrices $A$ and $B$. Now, using (\ref{Hmeas}),
we can perform integration over the angular variables and rewrite (\ref{PKH}) as
\be
e^{\frac{1}{2g}\tr L^2} \int_{n\times n}dh
\exp \left( -\frac{g}{2}\tr h^2
 +N\tr\log h-i\tr h L\right)\sim \\\sim
 e^{\frac{1}{2g}\sum_i L_i^2} \int \prod_idh_i
 {\Delta (h)\over\Delta (L)}\exp \left( -\frac{g}{2}\sum_i h_i^2
 +N\sum_i\log h_i-i\sum_i h_i L_i\right)=\\=e^{\frac{1}{2g}\sum_i L_i^2}\int \prod_idh_i
 {\det h_i^{j-1}\over\Delta (L)}\exp \left( -\frac{g}{2}\sum_i h_i^2
 +N\sum_i\log h_i-i\sum_i h_i L_i\right)={\det \Phi_i(L_j)\over\Delta (L)}
\ee
where
\be
\Phi_i(L)\equiv e^{\frac{1}{2g} L^2}\int dx x^{i-1}\exp \left( -\frac{g}{2}x^2
 +N\log x-ixL\right)
\ee
It remains to note that the orthogonal polynomials with the measure
$\exp\left(-{1\over 2g}H^2\right)$ are the Hermit polynomials which have the integral
representation
\be
\tilde P_k(x)={g^{k/2}\over \sqrt{2\pi}}\int_{-\infty}^{+\infty}dy\left({x\over\sqrt{g}}+iy\right)^k
e^{-y^2/2}={i^ng^{k+{1\over 2}}\over\sqrt{2\pi }}\ e^{{x^2\over 2g}}\int_{-\infty}^{+\infty}dyy^k
e^{-{gy^2\over 2}-ixy}
\ee
This finally reduces (\ref{DMM}) to (\ref{PKH}).

\subsubsection{Proof III: Faddeev-Popov trick}

Another way to connect the two matrix integrals, suggested recently in
\cite{ShaW} is by using the Faddeev-Popov trick. In order not to make calculations with
Grassmann variables, we choose the opposite sign in (\ref{HMit}), (\ref{HMit0})
\be
t_k = \frac{g}{k}\tr L^{-k}, \ \ \ k\geq 1
\label{HMit-}
\ee
\be\label{HMit0-}
t_0=-g \ \tr \log L
\ee
 Then,
(\ref{PKH}) should be substituted with
(the results for the two choices of sign can be also related by continuation)
\be
\sim e^{-\frac{1}{2g}\tr L^2} \int_{n\times n}
\exp \left( -\frac{g}{2}\tr h^2
 -N\tr\log h+\tr h L\right)
dh
\label{PKH-}
\ee

\bigskip

If Miwa transform (\ref{HMit-}) is made in the original integral
(\ref{Horin}), it becomes \cite{UFN3} \be (\ref{Horin}) =  (\det
L)^{-N}\int_{N\times N} \frac{e^{-\frac{1}{2g} \Tr H^2} dH}
{\det(I\otimes I -H \otimes L^{-1})} = \int\int\int e^{-\frac{1}{2g}
\Tr H^2 + B\big(I\otimes L - H\otimes I\big)C} dH d^2B,
\label{Horintrint} \ee where Faddeev-Popov trick is applied to
substitute the determinant in the denominator by an integral over
auxiliary rectangular $N\times n$ complex matrix fields $B$ and
$C=B^\dagger$. Here $d^2B = dBdC = \prod_{i=1}^N\prod_{a=1}^n
d^2B_{ia}$ and \be B\big(H\otimes I - I\otimes L\big)C =
B_{ai}H_{ij}C_{ja} - B_{ia}L_{ab}C_{bi} = \tr B^\dagger H B - \Tr
BLB^\dagger \ee Taking the Gaussian integral over $H$ we finally
obtain: \be (\ref{Horin})= g^\frac{N^2}{2}\int\int e^{\frac{g}{2}\Tr
BB^\dagger BB^\dagger + \Tr BLB^\dagger}d^2B \label{Horindint} \ee
At the same time the Kontsevich-Penner integral (\ref{PKH-})
is equal to
\be
(\ref{PKH-}) = g^{\frac{N^2}{2}+\frac{n^2}{2}}\int \frac{e^{-\frac{g}{2}\tr h^2}dh}
{\det\left(L+g h\right)^N} =
g^{\frac{N^2}{2}+\frac{n^2}{2}}\int \frac{e^{-\frac{g}{2}\tr h^2}dh}
{\det(I\otimes g h+I\otimes L)} =\nn\\
=g^{\frac{N^2}{2}+\frac{n^2}{2}}\int\int\int e^{-\frac{g}{2} \tr h^2 +
B\big(I\otimes g h + I\otimes L\big)C}
dh dB dC
\ee
where the fields $B,C$ are exactly the same as in
(\ref{Horintrint}), while
\be
B\big(I\otimes g h + I\otimes L\big)C =
g B_{ia}h_{ab}C_{bi} + B_{ia}L_{ab}C_{bi} =
g \Tr B hB^\dagger   + \Tr BLB^\dagger
\ee
Again we can take the Gaussian integral over $h$ and obtain:
\be
(\ref{PKH-})=
g^{\frac{N^2}{2}}\int\int e^{\frac{g}{2}\Tr BB^\dagger BB^\dagger  +
\Tr BLB^\dagger}d^2B
\ee
i.e. exactly the same expression as at the r.h.s. of
(\ref{Horindint}).
Thus we conclude that
\be
(\ref{Horin}) = (\ref{PKH-})
\ee
the two integral representation for $Z_H(t)$ coincide.

Inverting the argument, the two matrix-integral
representations (\ref{Horin}) and (\ref{PKH-}) for $Z_H(t)$
are associated with two ways to decompose the quartic vertex
$\Tr BB^\dagger BB^\dagger = \tr B^\dagger BB^\dagger B$
with the help of auxiliary fields $H$ and $h$,  coupled
respectively to $BB^\dagger$ and $B^\dagger B$ and thus having
the different sizes: $N\times N$ and $n\times n$.

As a word of precaution we remind only that for any finite $n$
the Miwa transform (\ref{HMit}) defines only an $n$-dimensional
{\it subset} in the infinite-dimensional space of $t$-variables:
when expressed through $L$
the higher $t_k$ with $k>n$ are actually algebraic functions
of the lowest $t_1,\ldots,t_n$. Thus $Z_H(t)$ in this context
should be interpreted as a projective limit at $n\rightarrow\infty$.

\subsection{Genus expansion and the first multiresolvents}

Multiresolvents for Hermitian model are described in detail in
the reference-paper \cite{spef}.
Here we remind only the simplest of the relevant formulas.

Multiresolvents are defined by eq.(\ref{muddef}) and the first
step is to rewrite Virasoro constraints as recurrent relations
for $\rho^{(p|q)}$.
Such recursive reformulation is possible only if the
\textit{genus expansion} of the \textit{free energy}
${\cal F} = \log Z$ is performed, ${\cal F} = \sum_{p=0}^\infty
g^{2p-2}{\cal F}^{(p)}$. As explained in some detail in
\cite{spef}, this requirement picks up some rather special solutions
to Virasoro constraints, and only such solutions possess well-defined
multiresolvents and are associated with the \textit{bare} spectral curves
of finite genera.\footnote{It deserves emphasizing that
\textit{genus} $p$ in "genus expansion" refers to the genus of
the fat-graph Feynman diagrams contributing to ${\cal F}^{(\!p\,)}$.
It has nothing to do with the genus of the bare spectral curve,
throughout this text \textit{this} genus will be only zero,
while the genus of the \textit{full} spectral curve
(which defines the point of the Universal Grassmannian \cite{UMS}
underlying the matrix-model $\tau$-function) is infinite. Relation
between bare and full spectral curves is rather tricky and is not
yet fully clarified in the literature.}
The bare spectral curve $\Sigma$ is defined from non-linear equation
for $\rho^{(0|1)}$ -- the starting point of the recursion.
The next step provides $\rho^{(0|2)}$, which appears to be easily connected with Bergman
kernel bi-differential on $\Sigma$ \cite{CMMV}.
At each step of recursion there exists a certain arbitrariness
in the choice of solutions, which is, however, absent for the Virasoro
constraint (\ref{Hdico}) -- crucial for this \textit{unambiguity} is the
form of the l.h.s. of (\ref{Hdico}): the fact that it is obtained
by the \textit{shift} of the time variable $t_2 \rightarrow t_2 - 1/2$.
\textit{Shift} is parameter which is allowed to stand in denominators
when we build up a formal-series solution to Virasoro constraints.
The shift of $t_2$ is obviously associated with the integral
(\ref{Horin}) and defines what is naturally called the \textit{Gaussian}
phase of Hermitian model.
If other time or many times are shifted, then arbitrariness is
unavoidable, see \cite{AMMcheck} for its full
description. If partly fixed and parameterized by several arbitrary
variables, it provides the family of Dijkgraaf-Vafa non-Gaussian
partition functions \cite{DV,DVfol,CMMV}.\footnote{Relation between generic
genus-expansion-possessing solutions of \cite{AMMcheck} and
the Dijkgraaf-Vafa family is very similar to that between the
"general" and "total" solutions to the Hamilton-Jacobi equation,
see \cite[sect.7]{LL1}.}

\paragraph{Gaussian phase.}
The bare spectral curve for Gaussian phase of Hermitian model is
\be
\Sigma_{HG}:\ \ y^2 = z^2-4S
\ee
where $S=gN$ and a few first Gaussian multiresolvents are:
\be
\rho_H^{(0|1)}=\frac{z-\sqrt{z^2-4S}}{2}
\ee
\be
\rho_H^{(0|2)}(z_1,z_2) = \frac{1}{2(z_1-z_2)^2}\left(
\frac{z_1z_2-4S}{y(z_1)y(z_2)} - 1\right)
\label{rho02}
\ee
\be
\rho_H^{(1|1)}(z) = \frac{S}{y^5(z)}
\ee
\be
\rho_H^{(0|3)}(z_1,z_2,z_3) =
\frac{2S (z_1z_2 + z_2z_3 + z_3z_1 + 4S)
}{y^3(z_1)y^3(z_2)y^3(z_3)}
\label{rho03}
\ee
\be
\rho_H^{(1|2)}(z_1,z_2) = \frac{S}{y^7(z_1)y^7(z_2)}
\Big(
z_1z_2(5z_1^4 + 4z_1^3z_2 + 3z_1^2z_2^2 + 4z_1z_2^3 + 5z_2^4) +
 \nn \\  +
4S\left(z_1^4 - 13z_1z_2(z_1^2 + z_1z_2 + z_2^2) + z_2^4\right)+
16S^2(-z_1^2 + 13z_1z_2 - z_2^2) + 320S^3
\Big)
\label{rho12}
\ee
\be
\rho_H^{(2|1)}(z) = \frac{21S\left(z^2 + S\right)}
{y^{11}(z)}
\ee
They are deduced from the recurrent relations
\be
y(z_1)\rho_H^{(0|2)}(z_1,z_2) =
\partial_{z_2}\frac{\rho_H^{(0|1)}(z_1)-\rho_H^{(0|1)}(z_2)}
{z_1-z_2}
\ee
\be
y(z_1)\rho_H^{(1|1)}(z_1) =
\rho_H^{(0|2)}(z_1,z_1)
\ee
\be
y(z_1)\rho_H^{(0|3)}(z_1,z_2,z_3) =
2\rho_H^{(0|2)}(z_1,z_2)\rho_H^{(0|2)}(z_1,z_3) + \nn \\ +
\partial_{z_2}\frac{\rho_H^{(0|2)}(z_1,z_3)-\rho_H^{(0|2)}(z_2,z_3)}
{z_1-z_2} +
\partial_{z_3}\frac{\rho_H^{(0|2)}(z_1,z_2)-\rho_H^{(0|2)}(z_2,z_3)}
{z_1-z_3}
\ee
\be
y(z_1)\rho_H^{(1|2)}(z_1,z_2) =
2\rho_H^{(0|2)}(z_1,z_2)\rho_H^{(1|1)}(z_1) +
\rho_H^{(0|3)}(z_1,z_1,z_2) +
\partial_{z_2}\frac{\rho_H^{(1|1)}(z_1)-\rho_H^{(1|1)}(z_2)}
{z_1-z_2}
\ee
\be
y(z_1)\rho_H^{(2|1)}(z_1) =
\left(\rho_H^{(1|1)}(z_1)\right)^2
+ \rho_H^{(1|2)}(z_1,z_1)
\ee

\paragraph{Non-Gaussian phases.}
For the sake of completeness we also give some formulas for
non-Gaussian phases. If instead of the Gaussian shift
$t_2\rightarrow t_2-\frac{1}{2}$ we apply
$t_k \rightarrow t_k - T_k$ with $W(z) = \sum_{k=0}^{n+1} T_kz^k$
and rewrite the shifted Virasoro constraints in terms of the
multiresolvents we get:
\be
W'(z)\rho(z) =\
\rho^2(z) + f(z) + g^2\hat\nabla(z)\rho(z) +
\hat P^-_z\left[v'(z)\rho(z)\right]
\label{virconfW}
\ee
where
\be
\hat\nabla(z)=\sum_{k=0}^\infty\frac{1}{z^{k+1}}\frac{\p}{\p t_k}
\ee
and
\be
\rho(z)=\hat \nabla(z) g^2\log Z
\ee
\be
f(z) = \hat P^+_z\left[W'(z)\rho(z)\right]
= \hat R_H(z) g^2\log Z
\ee
\be
W'(z)\rho_W^{(p|m+1)}(z,z_1,\ldots,z_m) -
f_W^{(p|m+1)}(z|z_1,\ldots,z_m) = \nn \\ =
\sum_q \sum_{m_1+m_2=m}
\rho_W^{(q|m_1+1)}(z,z_{i_1},\ldots,z_{i_{m_1}})
\rho_W^{(p-q|m_2+1)}(z,z_{j_1},\ldots,z_{j_{m_2}}) + \nn \\ +
\sum_{i=1}^m \frac{\partial}{\partial z_i}
\frac{\rho_W^{(p|m)}(z,z_1,\ldots,\check z_i,\ldots,z_m) - \rho_W^{(p|m)}(z_1,\ldots,z_m)}{z-z_i} +
\hat\nabla(z)\rho_W^{(p-1|m+1)}(z,z_1,\ldots,z_m)
\label{recrel1}
\ee
\be
f_H^{(p|m+1)}(z|z_1,\ldots,z_m)=\check R_H(z)\rho^{(p|m)}(z_1,\ldots,z_m)
\ee
\be\check R_H(z)=P_z^+\left[W'(z)\check \nabla(z)\right]=-\sum_{a=0}^{n-1}\sum_{b=0}^{n-a-1}(a+b+2)T_{a+b+2}z^a\frac{\p}{\p T_b}
\ee
\bigskip

For further details we refer to \cite{spef,AMMcheck}
and references therein.

\pagebreak

\section{Complex-matrix model $Z_C(t)$}

\subsection{Original integral representation}

Complex matrix model was originally defined as an integral
over $N\times N$ complex matrices $\Phi$
\be
Z_{C}(t) = \int_{N\times N}  \exp \left( - \frac{\Tr \Phi\Phi^\dagger}{g}
+\sum_{k=0}^\infty \frac{t_k}{g}\Tr (\Phi\Phi^\dagger)^k\right) d^2\Phi
\label{Corin}
\ee
where
$d^2\Phi =\left(\frac{i}{2}\right)^{N^2} \prod_{i,j=1}^N d^2\Phi_{ij} =
\prod_{i,j=1}^N d{\rm Re}(\Phi_{ij})d{\rm Im}(\Phi_{ij}) =
\prod_{i,j=1}^N \frac{i}{2}d\Phi_{ij}d\Phi^\dagger_{ij}$.

\subsection{Eigenvalue representation}

One can express a complex matrix $\Phi$ through
Hermitian $H$ and unitary $U$ matrices,
\be
\Phi = UH,\ \ \ \ \ \Phi^\dagger = HU^\dagger
\label{PHU}
\ee
and, further, through diagonal matrix $D$ and two unitary matrices
$U$ and $V$:
\be
\Phi = UDV^\dagger,\ \ \ \ \ \Phi^\dagger = VDU^\dagger
\ee

The norm of $\Phi$ decomposes as
\be
\Tr \delta\Phi \delta\Phi^\dagger = \Tr (\delta H)^2
-\Tr H^2(U^\dagger \delta U)^2 + \Tr [H,\delta H] (U^\dagger \delta U) =
\nn \\ =
\Tr (\delta D)^2 -\Tr (U^\dagger \delta U)^2 D^2
- \Tr D^2(V^\dagger \delta V)^2
+ 2\Tr (U^\dagger \delta U) D (V^\dagger \delta V) D,
\ee
so that the measure
\be\label{meC}
d^2\Phi = [dU][dV] \prod_{i<j}^N (H_i^2-H_j^2)^2 \prod_{i=1}^N H_idH_i
\ee
where $H_i$ are the eigenvalues of matrix $H$ (i.e. the entries of $D$).
In particular, for $N=1$ we have $\Phi = e^{i\theta} H$ and
$d^2\Phi = HdHd\theta$.

Comparing with (\ref{Hmeas0}), (\ref{Hmeas}), one can
see that for complex matrices
the measure is actually the same as for Hermitian matrix $H^2$,
such that $\Phi\Phi^\dagger = UH^2U^\dagger$.
Since action in the model (\ref{Corin}) also respects this substitution,
we obtain:
\be
Z_{C}(t) = V_N^2\int_0^\infty \prod_i dH^2_i\Delta^2(H^2)\exp\left(
-{1\over g}\sum_i H_i^2+{1\over g}\sum_{i,k}t_kH^{2k}_i
\right)\sim\nn\\\sim
\int [dU] \int \exp
\left( -{1\over g}\Tr H^2 + {1\over g}\sum_{k=0}^\infty t_k\Tr H^{2k}\right) d(H^2)
\label{CMHU}
\ee

This integral looks just the same as (\ref{Horin}) for Hermitian matrix
model, however, there is a difference in the integration contour:
$H^2$ is not an arbitrary Hermitian matrix. Relation between (\ref{Horin})
and (\ref{CMHU}) is like between an integral over entire real axis and
over its positive ray: the answers are different and even invariance
properties -- and thus the Picard-Fucks equations (Ward identities) are not
exactly the same.

\subsection{Virasoro constraints}

Since the
Eq.(\ref{CMHU}) means that the  Virasoro constraints are the same as
the "discrete Virasoro constraints" for Hermitian matrix model.
However, there are two differences.

First, $L_{-1}$-constraint, associated with the shift
$\delta(H^2) = \epsilon$, is excluded, because it would
correspond to a singular transform $\delta H \sim H^{-1}$.
This exclusion can also be considered as a result of the
above-mentioned change of
integration contour: from entire real line in the case
of Hermitian model to a positive ray  $0 \leq H^2 < \infty$
in the case of (\ref{CMHU}).

Second, the shift of time variables is
$t_k = \tilde t_k- \frac{1}{2}\delta_{k,2}$
in the Gaussian phase of Hermitian model, but it is rather
$t_k = \tilde t_k-\delta_{k,1}$
in the Gaussian phase of (\ref{CMHU}).

The two changes together make the seemingly diminished
set of Virasoro constraints
\be
\hat L_m Z_C = 0,\ \ \ m\geq 0,
\label{Cvico}
\ee
with
\be
L_m(t) = -\frac{\partial}{\partial t_{1+m}} +
\sum_{k\geq 1}^\infty kt_k \frac{\partial}{\partial t_{k+m}}
+ g^2 \sum_{a+b=m}\frac{\partial^2}{\partial t_a\partial t_b}
\ee
and
\be
\frac{\partial Z_C}{\partial t_0} = \frac{S}{g^2}Z_C
\ee
fully exhaustive: despite the lack of $\hat L_{-1}$,
these equations are enough for {\it unambiguous} recursive
reconstruction of all terms in the formal series
$Z_C(t)$ for the Gaussian branch of the complex matrix
model.

\subsection{Determinant representation and integrability}

The integrable properties of the complex matrix model are practically identical to those of the
Hermitian model. In particular, the partition function
\be \label{detrepTC3} Z_C \sim {\cal Z}^C_N=\det
\partial ^{i+j-2} C
\ee
where the moment matrix is given a bit different integral as compared with the Hermitian case,
\be
C_{ij}\equiv\int_0^\infty dx\exp\left(-{1\over g}x+{1\over g}\sum_kt_kx^k\right)
\ee
This is still a Toda chain $\tau$-function, however, it corresponds to another solution to the
hierarchy given by the Virasoro constraints (\ref{Cvico}).

\subsection{Kontsevich-Penner representation \label{Cpen}}

Like in the case of Hermitian model, the set of constraints
(\ref{Cvico}) has another matrix-integral solution \cite{compen},
different from (\ref{Corin}):
\be
Z_C(t) = \pi^{N^2-n^2}e^{-\tr \eta\eta^\dagger}
\int \exp\left(-\tr \phi\phi^\dagger +
\tr \eta^\dagger \phi + \tr \phi^\dagger \eta
+ N \tr\log (\phi\phi^\dagger)\right)  d^2\phi
\label{PKC}
\ee
This time integral is over complex matrices $\phi$,
but their size $n$ is, like in (\ref{PKH}), independent of $N$,
which appears only as a parameter in the Penner term. For the sake of simplicity,
we put here $g=1$, the $g$-dependence being easily restorable. The time variables are
related to the external matrices $\eta$ and $\eta^\dag$ as
\be t_k =-
\frac{1}{k}\tr(\eta\eta^\dagger)^{-k}, \ \ \ \ k\geq 1, \\
t_0=\log(\eta\eta^\dag)\label{CMit}
\ee

\subsubsection{Proof I: Ward identities}

The Ward identity associated with the shifts $\phi \rightarrow \phi +
\epsilon$ of the integration variable, is now
\be
\left[-{\p\over\p\eta_{tr}}+N\left({\p\over\p\eta_{tr}^\dag}\right)^{-1}+\eta^\dag
\right]\int \exp\left(-\tr \phi\phi^\dagger +
\tr \eta^\dagger \phi + \tr \phi^\dagger \eta
+ N \tr\log (\phi\phi^\dagger)\right)  d^2\phi=0
\ee
Therefore, one gets
\be
\left[-\left({\p\over\p\eta_{tr}^\dag}+\eta\right)
\left({\p\over\p\eta_{tr}}+\eta^\dag\right)+N+\left({\p\over\p\eta_{tr}^\dag}+\eta\right)
\eta^\dag
\right]Z_C=
\left[-{\partial^2\over\partial\eta\partial\eta^\dag}+N-
{1\over 2}\eta{\partial\over\partial\eta}-{1\over 2}\eta^\dag{\partial\over
\partial\eta^\dag}\right]Z_C=0
\ee
and,
substituting $Z_C$ as a function of the Miwa variables (\ref{CMit})
one reproduces (\ref{Cvico}), see \cite{compen}.

\subsubsection{Proof II: Orthogonal polynomials}

Let us put $L\equiv \eta\eta^\dag$. Then, one can immediately repeat the calculation of
s.\ref{orthopol}
in order to obtain that (\ref{Corin}) is equal to (\ref{DMM}) and (\ref{v3.24}), where
the polynomials $\tilde P_k(H)$ are now orthogonal with the weight $\exp (-x)$ on the positive
real semi-axis. Such orthogonal polynomials are nothing but the Laguerre polynomials \cite{BE},
which have the following integral representation
\be\label{PC}
\tilde P_k(x)=(-1)^ke^x\int_0^\infty dy y^k e^{-y}J_0(2\sqrt{xy})
\ee
where $J_0(x)$ is the zero order Bessel function.

Now let us rewrite (\ref{PKC}) in the determinant form. This time in order to integrate over
angular variables, we need to use instead of the Itzykson-Zuber formula the following very nice
formula of integration over two unitary $n\times n$ matrices, \cite{BK}
\be
\int_{U(n)}dU\int_{U(n)}dV \exp\left({1\over 2}\tr\left[UAVB+B^\dag V^\dag A^\dag U^\dag
\right]\right)=2^{n(n-1)}V_n^2{\det J_0(x_iy_j)\over\Delta (x^2)\Delta (y^2)}
\ee
where $x_i^2$ and $y_j^2$ are the eigenvalues of $A^\dag A$ and $BB^\dag$ respectively, $A$
and $B$ being arbitrary $n\times n$ complex matrices. Using this formula and (\ref{meC})
and denoting eigenvalues of $\phi\phi^\dag$ and $\eta\eta^\dag$ through $y_i$ and
$x_i$ respectively, one obtains
\be
e^{-\tr \eta\eta^\dagger}
\int \exp\left(-\tr \phi\phi^\dagger +
\tr \eta^\dagger \phi + \tr \phi^\dagger \eta
+ N \tr\log (\phi\phi^\dagger)\right)  d^2\phi\sim \\\sim
 e^{-\sum_i x_i} \int_0^\infty \prod_idy_i
 {\Delta (y)\over\Delta (x)}\exp \left( -\sum_i y_i
 +N\sum_i\log y_i\right)J_0(2\sqrt{x_iy_i})={\det \Phi^{(C)}_i(x_j)\over\Delta (x)}
\ee
where
\be
\Phi^{(C)}_i(x)=e^{-x}\int_0^\infty dy y^{i-1}\exp\left(
-y +N\log yJ_0(2\sqrt{xy})
\right)
\ee
Comparing this with (\ref{PC}) we ultimately identify (\ref{Corin}) and (\ref{PKC}).

\subsubsection{Proof III: Faddeev-Popov trick}

Direct equivalence of the two integrals (\ref{Corin}) and
(\ref{PKC}) can be proved by a somewhat tricky generalization
of Faddeev-Popov argument from \cite{ShaW}, which we applied
in s.\ref{Hpen} above.

As before, we make the other choice of the sign in the Miwa transform,
\be t_k =
\frac{1}{k}(\eta\eta^\dagger)^{-k}, \ \ \ \ k\geq 1
\ee
in order to deal with bosonic auxiliary fields.
After the Miwa transform, the integral (\ref{Corin})
becomes
\be
(\ref{Corin})
 = \int  \exp\left(-\Tr\Phi\Phi^\dagger + \sum_{k=0}^\infty
\frac{1}{k} \tr (\eta\eta^\dagger)^{-k}
\Tr (\Phi\Phi^\dagger)^k\right) d^2\Phi =
\det(- \eta\eta^\dagger)^N
\int  \frac{e^{-\Tr\Phi\Phi^\dagger}d^2\Phi}
{\det (\Phi\Phi^\dagger \otimes I - I \otimes \eta\eta^\dagger)}
= \nn \\
= \det(- \eta\eta^\dagger)^N
\int\int e^{-\Tr\Phi\Phi^\dagger
+ \tr B^\dagger \Phi\Phi^\dagger B - \Tr
B\eta\eta^\dagger B^\dagger}
d^2\Phi d^2B
= \det(- \eta\eta^\dagger)^N
\int \frac{e^{-\tr\eta\eta^\dagger B^\dagger B}d^2B}
{\det_{N\times N} (I - BB^\dagger)^N}
\ee
Note that the last
determinant is raised to the power $N$, this is because we
integrate over $N^2$ complex-valued variables $\Phi_{ij}$:
the relevant piece of the action is $\sum_{i,j,k=1}^N
\Phi_{ij}\bar\Phi_{ik}
\left(\delta_{jk} - \sum_{a=1}^n B_{ja}\bar B_{ka}\right)$.

At the same time integral (\ref{PKC}) is:
\be
(\ref{PKC}) = \int \frac{e^{-\tr\phi\phi^\dagger}d^2\phi}
{\det_{n\times n}^N\!
\Big((\phi + \eta)(\phi^\dagger + \eta^\dagger)\Big)}
= \int\int e^{-\tr\phi\phi^\dagger
- \Tr b^\dagger(\phi-\eta)(\phi^\dagger-\eta^\dagger)b}d^2\phi d^2b
= \int \frac{e^{-\tr \eta\eta^\dagger \frac{b^\dagger b}{1+b^\dagger b}}
d^2b}{\det_{n\times n} (1+b^\dagger b)^n}
\ee
Like $B$, $b$ is rectangular $N\times n$ matrix, but the integrals
are not literally equal as it was in the case of Hermitian model:
one still needs to relate $B$ and $b$.

Let us begin with a few examples.

\bigskip

{\bf Examples.} \underline{$N=n=1$}:
In this case we can introduce new variables:
$\rho = B^\dagger B = |B|^2$ and $\sigma = b^\dagger b = |b|^2$.
Denoting also $K = \eta\eta^\dagger = |\eta|^2$, we obtain our
two integrals in the form:
\be
N=n=1:\ \ \ \
(\ref{Corin}) = \int \frac{e^{-\rho K}d\rho}{1-\rho}, \ \ \ {\rm while}
 \ \ \
(\ref{PKC}) = \int\frac {e^{-\frac{\sigma}{1+\sigma}K}d\sigma}
{1+\sigma}
\ee
and integrands coincide because for $\rho = \frac{\sigma}{1+\sigma}$
we have $d\rho = -\frac{d\sigma}{(1+\sigma)^2}$,
while $1-\rho = \frac{1}{1+\sigma}$.

\underline{$n=1$, $N$ arbitrary}:
In this case $B$ is a complex $N$-vector
$(B_1,\ldots,B_N)$, the $N\times N$ matrix
$BB^\dagger$ has rank $1$ and $\det(I-BB^\dagger) =
1- |B_1|^2-\ldots-|B_n|^2 = 1-\rho_1-\ldots-\rho_n = 1 -\rho_+$,
so that
\be
(\ref{Corin}) = \int \frac{e^{-\rho_+ K}d\rho_1\ldots d\rho_N}
{(1-\rho_+)^N} \sim \int \frac{\rho_+^{N-1}e^{-\rho_+ K}d\rho_+}
{(1-\rho_+)^N}
\ee
and similarly
\be
(\ref{PKC}) = \int\frac {e^{-\frac{\sigma_+}{1+\sigma}K}
d\sigma_1\ldots d\sigma_N}{1+\sigma_+}
\sim \int\frac {\sigma_+^{N-1}e^{-\frac{\sigma_+}{1+\sigma}K}
d\sigma_+}{1+\sigma_+},
\ee
where we used the fact that the volume of a simplex
$\rho_1>0, \ldots, \rho_N>0,\ \rho_1+\ldots +\rho_N = \rho_+$
is proportional to $\rho_+^{N-1}$ and similarly for $\sigma$'s.
Making the same transformation as above,
$\rho_+ = \frac{\sigma_+}{1+\sigma_+}$, we see again that the
integrals coincide.

\underline{$N=1$, $n$ arbitrary}:
This time $B$ and $b$ are complex $n$-vectors. If we perform an $SU(n)$
rotation to diagonalize $\eta\eta^\dagger \rightarrow
{\rm diag}(K_1,\ldots,K_n)$, then both integrals (\ref{Corin})
and (\ref{PKC}) still contain $B$ and $b$ only in the form
of squared modules $\rho_a = |B_a|^2$ and $\sigma_a = |b_a|^2$:
\be
(\ref{Corin}) = \int \frac{e^{-\sum_{a=1}^n\rho_a K_a}
d\rho_1\ldots d\rho_n}{1-\rho_+} \ \ \ \ \ {\rm and} \ \ \ \ \
(\ref{PKC}) = \int \frac{e^{-\sum_{a=1}^n\sigma_a K_a/(1+\sigma_+)}
d\sigma_1\ldots d\sigma_n}{(1+\sigma_+)^n}
\ee
The integrals are related by our usual change of variables
$\rho_a = \frac{\sigma_a}{1+\sigma_+}$, only the measure transform
gets a little trickier:
\be
\wedge_{a=1}^n d\rho_a = \wedge_{a=1}^n
\left(\frac{d\sigma_a}{1+\sigma_+}
- \frac{\sigma_a d\sigma_+}{(1+\sigma_+)^2}\right) =
\frac{\wedge_{a=1}^n d\sigma_a}{(1+\sigma_+)^n} -
\frac{\sum_{a=1}^n\sigma_a \wedge_{a=1}^n d\sigma_a}
{(1+\sigma_+)^{n+1}} =
\frac{1}{(1+\sigma_+)^{n+1}}\wedge_{a=1}^n d\sigma_a
\ee
Substituting also  $1-\rho_+ = (1-\sigma_+)^{-1}$,
we see that the two integrals are in fact the same.

\subsection{Genus expansions and the first multiresolvents}

\paragraph{Gaussian phase.}
The first resolvent \cite{spef} in this case is
\be
\rho^{(0|1)}_C(z) = \frac{1}{2}\left(1-\sqrt{1-\frac{4S}{z}}\right)
\ee
and the bare spectral curve
\be
\Sigma_C:\ \ \   y^2 = z(z-4S)
\ee
A few next multiresolvents are:
\be
\rho^{(0|2)}_C(z_1,z_2)=\frac{1}{2(z_1-z_2)^2}\left(
\frac{z_1z_2-2S(z_1+z_2)}{y_1y_2}-1\right)
\label{rho2C}
\ee
\be
\rho^{(1|1)}_C(z)=\frac{zS^2}{y_C(z)^5}
\ee
\be
\rho^{(0|3)}_C(z_1,z_2,z_3)=2\,\frac
{z_1z_2z_3{S}^{2}}{ y_C(z_1)^3y_C(z_2)^3y_C(z_3)^3}
\ee
\be
\rho_C^{(1|2)}(z_1,z_2)=\frac{z_1^2z_2^2S^2}{y_1^7y_2^7}\left(3z_2z_1^3+3z_2^3z_1
+2z_2^2z_1^2\right.\nn\\
\left.-2(z_1^3+17z_1^2z_2+17z_1z_2^2+z_2^3)S
+8(3z_1^2+19z_1z_2+3z_2^2)S^2-96(z_1+z_2)S^3+128S^4 \right)
\ee
\be
\rho_C^{(2|1)}(z)=\frac{(9S^2-8zS+8z^2)S^2z^3}{y(z)^{11}}
\ee

They are deduced from the recurrent relations
\be
\frac{y(z_1)}{z_1}\rho_C^{(0|2)}(z_1,z_2) =
\frac{1}{z_1}\partial_{z_2}\frac{z_1\rho_C^{(0|1)}(z_1)-z_2\rho_C^{(0|1)}(z_2)}
{z_1-z_2}
\ee
\be
\frac{y(z)}{z}\rho_C^{(1|1)}(z) = \rho_C^{(0|2)}(z,z)
\ee
\be
\frac{y(z_1)}{z_1}\rho_C^{(0|3)}(z_1,z_2,z_3) =
2\rho_C^{(0|2)}(z_1,z_2)\rho_C^{(0|2)}(z_1,z_3) + \nn \\ +
\frac{1}{z_1}\partial_{z_2}\frac{z_1\rho_C^{(0|2)}(z_1,z_3)-z_2\rho_C^{(0|2)}(z_2,z_3)}
{z_1-z_2} +
\frac{1}{z_1}\partial_{z_3}\frac{z_1\rho_C^{(0|2)}(z_1,z_2)-z_3\rho_C^{(0|2)}(z_2,z_3)}
{z_1-z_3}
\ee
\be
\frac{y(z_1)}{z_1}\rho_C^{(1|2)}(z_1,z_2) =
2\rho_C^{(0|2)}(z_1,z_2)\rho_C^{(1|1)}(z_1) +
\rho_C^{(0|3)}(z_1,z_1,z_2) +
\frac{1}{z_1}\partial_{z_2}\frac{z_1\rho_C^{(1|1)}(z_1)-z_2\rho_C^{(1|1)}(z_2)}
{z_1-z_2}
\ee
\be
\frac{y(z_1)}{z_1}\rho_C^{(2|1)}(z_1) =
\left(\rho_C^{(1|1)}(z_1)\right)^2 + \rho_C^{(1|2)}(z_1,z_1)
\ee

\paragraph{Non-Gaussian phases.}
A generic phase of the complex matrix model is given by first several time variables shifted,
$t_k\to T_k+t_k$. Then, for a generic polynomial potential $W(z)=\sum_{k=0}^{n+1} T_k z^k$
the Virasoro constraints for the complex-matrix model look like
\be
W'(z)\rho(z)=\rho^2(z)+f(z)+g^2\hat\nabla(z)\rho(z)+\frac{1}{z}P^-_z\left[z v'(z)\rho(z)\right]
\ee
where
\be
f(z)=\frac{1}{z}P^+_z\left[z W'(z)\rho(z)\right]
\ee
Higher resolvents can be extracted from the equations
\be
W'(z)\rho_{W}^{(p|m+1)}(z,z_1,\ldots,z_m) -
f_{C}^{(p|m+1)}(z|z_1,\ldots,z_m) = \nn \\
= \sum_q
\sum_{m_1+m_2=m}
\rho_{W}^{(q|m_1+1)}(z,z_{i_1},\ldots,z_{i_{m_1}})
\rho_{W}^{(p-q|m_2+1)}(z,z_{j_1},\ldots,z_{j_{m_2}}) +\nn\\+
\sum_{i=1}^m\frac{1}{z} \frac{\partial}{\partial z_i}
\frac{z\rho_{W}^{(p|m)}(z,z_1,\ldots,\check z_i,\ldots,z_m) -
z_1\rho_W^{(p|m)}(z_1,\ldots,z_m)}{z-z_i} +
\hat\nabla(z)\rho_{W}^{(p-1|m+1)}(z,z_1,\ldots,z_m).
\ee
where
\be
f_C^{(p|m+1)}(z|z_1,\ldots,z_m)=\check R_C(z)\rho^{(p|m)}(z_1,\ldots,z_m)
\ee
with
\be
\check R_C(z)=\frac{1}{z}P_z^+\left[z W'(z)\check \nabla(z)\right]=-\sum_{a=-1}^{n-1}\sum_{b=0}^{n-a-1}(a+b+2)T_{a+b+2}z^a\frac{\p}{\p T_b}
\ee
Quadratic equation for simplest resolvent leads to the answer
\be
\rho^{(0|1)}_C(z)=\frac{W'(z)-\frac{y_C(z)}{z}}{2}
\ee
where
\be
y_C(z)^2=z\left(W'^2-4\check R_C(z) F^{(0)}_C\right)
\ee
For the Gaussian complex model $T_k=\delta_{k,1}$, so $W(z)=z$, $\check R_C(z)=-\frac{1}{z}\frac{\p}{\p T_0}$

\pagebreak

\section{Kontsevich model $Z_K(\tau)$}

\subsection{Original integral representation}

Kontsevich model was originally defined in \cite{Ko} as a generating
function of topological indices
of the moduli space of Riemann surfaces.
M.Kontsevich represented this generating function in form of the
now-famous matrix integral over auxiliary $n\times n$ dimensional
Hermitian matrices (one can easily introduce into this integral the parameter $g$
similarly to (\ref{PKH}), \cite{AMMIM} however, for the sake of simplicity, we put here $g=1$):
\be
 Z_K=
\exp\left(-{2\over 3}\tr L^3\right){\int_{n\times n} dh\
\exp\left(-{1\over 3}\tr h^3+\tr L^2 h\right)\over
\int_{n\times n} dh\ \exp\left(-\tr Lh^2\right)}
\label{Korin}
\ee
where time-variables are Miwa-transformed:
\be
\tau_k = \frac{1}{k}\tr L^{-k}
\label{TL}
\ee
If expressed through the $\tau$-variables, $Z_K(\tau)$ is actually independent
of auxiliary parameter $n$.
This model can be further generalized to Generalized Kontsevich Model (GKM) \cite{GKM},
\be\label{GKM}
Z_{GKM}=  \frac{\int_{n\times n} e^{-U(L,h)}dh}
{\int_{n\times n} e^{-U_2(L,h)}dh}
\ee
where
\be
U(L,h) \equiv \tr[{\cal V}(L+h) - {\cal V}(L) - {\cal V}'(L)h]
\ee
and
\be
U_2(L,h) =   \lim_{\epsilon \rightarrow 0}
{1\over \epsilon ^2} U(L,\epsilon h)
\ee
is an $h^2$-term in $U$. $W$ is here an arbitrary (power series) potential and $Z_{GKM}$ is
a function of the same
($W$-independent) Miwa transform (\ref{TL}).
Many properties of GKM are in fact independent of the choice of ${\cal V}(h)$.

In fact, one can also consider the matrix model (\ref{GKM}) with a different normalization
as a function of time variables
\be
{\cal T}_k={1\over k}\tr \Lambda^k
\ee
where the matrix $\Lambda={\cal V}'(L)$ enters in {\it positive} powers.
This function is called the character phase and is considered in detail in
\cite{GKMU}. In this paper we restrict ourselves with the Kontsevich phase only, where the time
variables are given by (\ref{TL}).

Note also that it is often convenient to fix ${\cal V}(h)$ to be a polynomial of $h$
(polynomial Kontsevich model, \cite{GKM}) or that of $h^{-1}$
(antipolynomial Kontsevich model \cite{GKMU}). In this section we consider only the polynomial
case, leaving the antipolynomial one until the next section
(where it emerges within the context of the unitary matrix model).

\subsection{Eigenvalue representation}

Shifting the integration variable $h\to h-L$ one obtains that
\be
Z_{GKM}\sim\int_{n\times n}dh \exp\left(-\tr {\cal V}(h)-\tr {\cal V}'(L)h\right)
\ee
Now using the Itzykson-Zuber formula and (\ref{Hmeas}), one can perform integration
over the angular variables in this integral:
\be
\int_{n\times n}dh \exp\left(-\tr {\cal V}(h)-\tr {\cal V}'(L)h\right)\sim
{\det_{ij}F_i(\lambda_j)\over\Delta (\lambda)}
\ee
where $\lambda_i$ are the eigenvalues of the matrix ${\cal V}'(L)$ and
\be
F_i(\lambda)=\int dxx^{i-1}\exp\left(-{\cal V}(x)+\lambda x\right)
\ee

\subsection{Virasoro constraints}

Straightforward Ward identities for $Z_K$ are as previously associated with the
shift $h \rightarrow h+\epsilon$ of integration variable $h$:
\be
\left[\left({\p\over\p L_{tr}^2}\right)^2-L^2\right]\int_{n\times n} dh\
\exp\left(-{1\over 3}\tr h^3+\tr L^2 h\right)=0
\ee
Now one should take into account the normalization factor and come to the $\tau$-variables.
Conversion to the $\tau$-variables is highly non-trivial, it
was first performed in \cite{grama} and leads to the celebrated result
\cite{FKN,DVV}:
\be\label{VirCon}
\hat {\cal L}_n Z_K = 0, \ \ \ \hat {\cal L}_n
= {1\over 2}\sum _{{k\geq \delta_{n+1,0} }\atop {k\ odd}}
k\tau_k {\partial \over \partial T_{k+2n}} + {1\over 4}
\sum _{^{a+b=2n}_{a,b\geq 0\ ;\ a,b\ odd}}
{\partial ^2\over \partial \tau_a\partial \tau_b} + \nn \\
+ \delta _{n+1,0}\cdot {\tau^2_1\over 4} + \delta _{n,0}\cdot {1\over 16}
- {\partial \over \partial \tau_{2n+3}}
\ee
This proved equivalence of Witten's topological $2d$ gravity
\cite{Witogra} to $Z_K$ and -- since $Z_K$ is trivially a KP
tau-function -- proved that partition function of $2d$ gravity is
indeed a tau-function (as anticipated in \cite{KazGro}). Analogous
conversion to $T$-variables of the Ward identities for $Z_{GKM}$ is
even more sophisticated and give rise to $W$-constraints (or $\tilde W$-constraints in
the character phase) \cite{tWco,GKMU}.

\subsection{Determinant representation and integrability
\label{Kdere}}

Now one can take into account all the normalization factors and
further transform this determinant (after quite tedious calculation, \cite{GKM}) to
\be\label{detGKM}
Z_{GKM}={\det_{ij}\phi_i(L_j)\over\Delta (L)}
\ee
where
\be
\phi_i(L)=\sqrt{{\cal V}''(L)}e^{{\cal V}(L)-L{\cal V}'(L)}F_i({\cal V}'(L))
\ee
Formula (\ref{detGKM}) if true for any number of
Miwa variables (size of the determinant) fixes a KP hierarchy $\tau$-function \cite{tauf,versus}
that depends on times
\be\label{MiwaT}
\tau_k={1\over k}\sum_i L_i^{-k}
\ee
provided the asymptotics of $\phi_i(L)$ at large $L$ is
\be\label{asym}
\phi_i(L)\stackrel{L\to\infty}{\sim} L^{i-1}\left(1+O(1/L)\right)
\ee
In particular, (\ref{asym}) guarantees that $Z_{GKM}$ is a function of variables $\tau_k$
(\ref{MiwaT}) and does not depend on their number.

Now if one takes the monomial potential ${\cal V}(h)=h^{p+1}$ (the case of a polynomial
potential of degree $p+1$ describes a hierarchy equivalent to the monomial case, see
details in \cite{LGM}), the partition function $Z_{GKM}$ is a $\tau$-function of the
$p$-reduced KdV hierarchy, which does not depend on times $\tau_{pk}$ for all $k$. In particular,
the Kontsevich partition function (\ref{Korin}) is a KdV $\tau$-function depending only on
odd times $\tau_{2k+1}$. A concrete solution of the KdV hierarchy is fixed by the Virasoro constraints
(\ref{VirCon}) (in fact, it is enough to use only the lowest constraint in addition to the
KdV hierarchy equations in order to fix the partition function unambiguously).

Note that one could starts from the Virasoro constraints (\ref{VirCon}) instead of the
matrix integral. Then, there are much more solutions, the KdV one corresponding only to
distinguished solutions of the Dijkgraaf-Vafa type \cite[2nd paper]{spef}.

Note that one can easily continue the (Generalized) Kontsevich matrix integral to the whole
Toda lattice hierarchy adding to $U(L,h)$ (but not to $U_2(L,h)$) the term
\be
\Delta U(L,h)=\Delta {\cal V}(L+h)-\Delta {\cal V}(L),\ \ \ \ \ \ \Delta {\cal V}(h)=
\aleph\log h-\sum_k \bar \tau_kh^{-k}
\ee
Here $\aleph$ is the zeroth (discrete) Toda time and $\bar \tau_k$ are the negative Toda times.
In the special case of quadratic potential ${\cal V}(h)=h^2$ this matrix integral reduces to the
Toda chain, as we observed in s.2.

\subsection{Kontsevich-Penner representation}

Of course, Kontsevich model is already in the Kontsevich form.
No parameter $N$ is obligatory present and no Penner term is needed (until one wants to
deal with the (Generalized) Kontsevich integral as with the
Toda lattice hierarchy).

\subsection{Genus expansion and the first multiresolvents}

\paragraph{Generic phase.}
Similarly to the Hermitian and complex matrix models, a generic phase of the Kontsevich model
is given by first several time variables shifted $\tau_{2k+1}\to\tau_{2k+1}+T_k$. In this case,
\be
\hat\nabla(z)=\sum_{k=0}^\infty \frac{1}{z^{k+1}}\frac{\p}{\p \tau_{2k+1}}
\ee
\be
W'(z)=\sum_{k=0}^{n+1}\left(k+\frac{1}{2}\right)T_k z^k
\ee
\be
v'(z)=\sum_{k=0}^{\infty}\left(k+\frac{1}{2}\right)\tau_{2k+1} z^k
\ee
\be
W'(z)\rho(z)=\rho^2(z)+f_K(z)+g^2\hat\nabla(z)\rho(z)+z P_z^-\left[\frac{v'(z)\rho(z)}{z}\right]+\frac{g^2}{16z}+\frac{(\tau_0-T_0)^2}{16}
\ee
\be
f_K(z)=zP^+_z\left[\frac{W'(z)\rho(z)}{z}\right]=g^2 \hat R_K(z)\log Z
\ee
\be
\hat R_K(z)=-\sum_{k=2}^{n+1}\sum_{m=0}^{k-2}\left(k+\frac{1}{2}\right)T_kz^{k-m-1}\frac{\p}{\p T_m}
\ee
\be
\rho(z)=g^2\hat\nabla(z)\log Z
\ee
\be
W'(z)\rho_W^{(p|m+1)}(z,z_1,\ldots,z_m) -
f_W^{(p|m+1)}(z|z_1,\ldots,z_m) = \nn \\ =
\sum_q \sum_{m_1+m_2=m}
\rho_W^{(q|m_1+1)}(z,z_{i_1},\ldots,z_{i_{m_1}})
\rho_W^{(p-q|m_2+1)}(z,z_{j_1},\ldots,z_{j_{m_2}}) + \nn \\ +
\sum_{i=1}^m \left(\frac{\partial}{\partial z_i}-\frac{1}{2z_i}\right)
\frac{z_i\rho_W^{(p|m)}(z,z_1,\ldots,\check z_i,\ldots,z_m) - z\rho_W^{(p|m)}(z_1,\ldots,z_m)}{z-z_i} +
\hat\nabla(z)\rho_W^{(p-1|m+1)}(z,z_1,\ldots,z_m)+\nn\\
\frac{\delta_{p,1}\delta_{m,0}}{16z}+\delta_{p,0}\left(\frac{\delta_{m,0}T_1^2}{16}-\frac{\delta_{m,1}T_1}{8z_1}+\frac{\delta_{m,2}}{8z_1z_2}\right)
\ee
\paragraph{Gaussian case.} In this case, $W'(z)=z+\frac{T_0}{2}$, $f^{(k|m)}=0$
\be
\rho^{(0|1)}(z)=\frac{z+\frac{T_0}{2}-\sqrt{z^2+T_0z}}{2}
\ee
\be
\rho^{(0|2)}(z_1,z_2)=\frac{1}{4(z_1-z_2)^2}\left(\frac{(z_1+z_2+2T_0)z_1z_2}{y(z_1)y(z_2)}-(z_1+z_2)\right)
\ee
\be
y(z_1)\rho^{(0|2)}(z_1,z_2)=\left(\frac{\p}{\p z_2}-\frac{1}{2z_2}\right)\frac{z_2\rho^{(0|1)}(z_1)-z_1\rho^{(0|1)}(z_2)}{z_1-z_2}-\frac{T_0}{8z_2}
\ee
\be
\rho^{(0|3)}(z_1,z_2,z_3)=\frac{z_1^2z_2^2z_3^2}{y(z_1)^3y(z_2)^3y(z_3)^3}
\ee
\be
y(z_1)\rho^{(0|3)}(z_1,z_2,z_3)=\left(\frac{\p}{\p z_2}-\frac{1}{2z_2}\right)\frac{z_2\rho^{(0|2)}(z_1,z_3)-z_1\rho^{(0|2)}(z_2,z_3)}{z_1-z_2}+\nn\\
\left(\frac{\p}{\p z_3}-\frac{1}{2z_3}\right)\frac{z_3\rho^{(0|2)}(z_1,z_2)-z_1\rho^{(0|2)}(z_2,z_3)}{z_1-z_3}+\frac{1}{8z_1z_2}
\ee
\be
y(z)\rho^{(1|1)}(z)=\rho^{(0|2)}(z,z)+\frac{1}{16z}
\ee
\be
\rho^{(1|1)}(z)=\frac{z^3}{16 y(z)^5}
\ee
\be
y(z_1)\rho^{(1|2)}(z_1,z_2)=2\rho^{(0|2)}(z_1,z_2)\rho^{(1|1)}(z_1)+\rho^{(0|3)}(z_1,z_1,z_2)+\left(\frac{\p}{\p z_2}-\frac{1}{2z_2}\right)
\frac{z_2\rho^{(1|1)}(z_1)-z_1\rho^{(1|1)}(z_2)}{z_1-z_2}
\ee
\be
y(z_1)\rho^{(2|1)}(z)=\left(\rho^{(1|1)}(z)\right)^2+\rho^{(1|2)}(z,z)
\ee
\be
\rho^{(2|1)}(z_1,z_2)=\frac{z_1^4z_2^4}{32}\frac{5(z_1+T_0)^2+3(z_1+T_0)(z_2+T_0)+5(z_2+T_0)^2}{y(z_1)^7y(z_2)^7}
\ee
\be
\rho^{(2|1)}(z)=\frac{105}{254}\frac{z^6}{y(z)^{11}}
\ee
\pagebreak

\section{BGW model $\tilde Z_K(\tau) = Z_{BGW}(\tau)$ }

\subsection{Original integral representation}

Brezin-Gross-Witten (BGW) model is defined as a generating
function for all correlators of unitary matrices with
Haar measure $[dU]$:
\be
Z_{BGW} = \int_{N\times N} [dU] \exp\left(\Tr J^\dagger U +
\Tr JU^\dagger\right)
\label{BGWorin}
\ee
The integral actually depends only on eigenvalues of Hermitian
matrix $M=JJ^\dagger$, i.e. on the time-variables of
the form $\tau_k = \Tr (JJ^\dagger)^k$.

Haar measure $[dU]$ for unitary matrices is non-linear,
it can be reduced to a flat measures in different ways.
One possibility is to express $U$ through Hermitian matrices,
$U=\frac{1+iH}{1-iH}$ \cite{UHaar},
which defines $[dU]$ as the flat Hermitian measure
$dH = \prod_{i,j=1}^N dH_{ij}$ with additional Jacobian factor,
$[dU] = {\cal J}(H)dH$, ${\cal J} = \det(1+H^2)^{-N}$.
Another possibility \cite{GKMU,ShaW} is to impose the
constraints on the complex matrices:
\be\label{trick}
[dU] = \int d^2\Phi \delta(\Phi\Phi^\dagger - I) =
\int_{N\times N} dh e^{-\Tr h} \int_{N\times N}
d^2\Phi e^{\Tr h\Phi\Phi^\dagger}
\ee
For certain actions the integral over $d\Phi$ can be explicitly
taken and this gives rise to reformulation of original
unitary-matrix model.

\subsection{Eigenvalue representation}

Since technically the most simple way to obtain eigenvalue representations is to start with
the Kontsevich-Penner representation of the BGW model, we first consider this representation.

\subsection{Kontsevich-Penner representation}

In variance with all other Kontsevich-Penner
representations, that of the BGW model connects the two integrals over the
two matrices (unitary and Hermitian ones) of the {\it same} size, \cite{GKMU}:
\be
Z_{BGW} = \int_{N\times N} dh \exp\Big(-\tr h^{-1} + \tr Mh - N\tr \log h\Big)
\label{PKBGW}
\ee
This makes theory of the BGW model somewhat harder and one sometimes
embeds it into the universal BGW model \cite{GKMU} with an arbitrary
coefficient in front of the logarithmic term. However, in order to make contact with the
BGW model (\ref{BGWorin}) one ultimately has to put this coefficient equal to $-N$.

\subsubsection{Proof I: Ward identities}

The simplest Ward identity for $Z_{BGW}$ has the form
\be
 \frac{\partial}{\partial J_{\rm tr}^\dagger}\cdot
\frac{\partial}{\partial J_{\rm tr}}
Z_{BGW}(J,J^\dagger) = I\cdot Z_{BGW}(J,J^\dagger).
\label{WI-BGWM-A}
\ee
or, equivalently, \cite{GKMU}
\be
\frac{\partial}{\partial M_{\rm tr}} M
  \frac{\partial}{\partial M_{\rm tr}} Z_{BGW}(M) = I\cdot Z_{BGW}(M).
\label{WI-BGWM-M}
\ee
At the same time, integral (\ref{PKBGW}) satisfies the equation
\be
\left[
\frac{\partial}{\partial M_{tr}} M
  \frac{\partial}{\partial M_{tr}} + (N-{\cal
N})\frac{\partial}{\partial M_{tr}} +
\left(\frac{\partial}{\partial M_{tr}}\right)^2 {\cal V}'
\left(\frac{\partial}{\partial M_{tr}}\right) \right]
\int_{N\times N} dh e^{{\tr}(Mh - {\cal N}\log h + {\cal V}(h))} = 0.
\label{WI-KI-BGWM}
\ee
At ${\cal N}=N$ and ${\cal V}'(h)=1/h$ (\ref{WI-BGWM-M}) and (\ref{WI-KI-BGWM}) coincide, which
establishes (\ref{PKBGW}).

\subsubsection{Proof II: Faddeev-Popov trick}

Another simple way to derive the Kontsevich-Penner representation for the BGW model is to use the
trick (\ref{trick}), \cite{ShaW}. Indeed,
\be
Z_{BGW} = \int dh e^{-\Tr h} \int d^2\Phi
\exp\left( \tr h\Phi\Phi^\dagger + \Tr J^\dagger\Phi
+ \tr J\Phi^\dagger\right) = \nn \\ =
\int_{N\times N} dh \exp\Big(-\tr h + \tr M/h - N\tr \log h\Big)\ \stackrel{h\rightarrow 1/h}{=}\\
\ \stackrel{h\rightarrow 1/h}{=}
\int_{N\times N} dh \exp\Big(-\tr h^{-1} + \tr Mh - N\tr \log h\Big)
\ee

The BGW model has two phases \cite{GKMU}: the \textit{Kontsevich phase},
where partition function is expanded in negative powers of $M$
and the \textit{character phase} where expansion goes in positive
powers of $M$. Below we describe them separately.

\subsection{On direct relation between $Z_C$ and $Z_{BGW}$}

In the Kontsevich-Penner form (\ref{PKC}) the complex matrix model
looks somewhat similar to original form (\ref{BGWorin}) of the BGW model. From (\ref{PKC}) one
obtains (representing $\phi=HU^\dag$ and $\phi^\dag=UH$):
\be
Z_C =\pi^{N^2-n^2}e^{-\tr\eta\eta^\dag} \int dH e^{-\tr H^2+2N\tr\log H}
Z_{BGW}(\eta^\dagger H^2\eta)
\ee
This tricky formula is the best direct relation known at present. A more transparent relation
is still lacking.

\subsection{Character phase}

In this phase the BGW partition function is considered as a function of the variables
\be\label{timesch}
{\cal T}_k={1\over k}\tr M^k
\ee
and one has to consider the Universal BGW model, i.e. the Kontsevich integral (\ref{PKBGW})
with an arbitrary coefficient of the logarithm, which is a free parameter and not the size
of the unitary matrix.

\subsubsection{Virasoro constraints}

Performing the change of variables in (\ref{WI-KI-BGWM}) from $M$ to ${\cal T}_k$, one can
directly obtain the Virasoro constraints satisfied by the BGW partition function $Z^+_{BGW}$
in the character phase:
\be
\hat L_m(N,{\cal T})Z^+_{BGW}=\delta_{m,1}Z^+_{BGW}, \ \ \ \ \ \ \ m\ge 1\\
\hat L_m(\alpha,{\cal T}) = \alpha\frac{\partial}{\partial {\cal T}_{m}} +
\sum_{k\geq 1}^\infty k{\cal T}_k \frac{\partial}{\partial {\cal T}_{k+m}}
+ \sum_{a+b=m}\frac{\partial^2}{\partial {\cal T}_a\partial {\cal T}_b}
\ee
Therefore, the Ward identity (and its solutions) depends on the size of matrix $N$. This means
that the integral (\ref{PKBGW}) is not just a function of variables ${\cal T}$, but also depends on
$N$. The way out is to consider the Universal BGW model given by the integral
\be
Z_{UBGW} = \int_{N\times N} dh \exp\Big(-\tr h^{-1} + \tr Mh - {\cal N}\tr \log h\Big)
\label{UBGW}
\ee
Then, the Virasoro constraints become
\be
\hat L_m(2N-{\cal N},{\cal T})Z^+_{UBGW}=\delta_{m,1}Z^+_{UBGW}, \ \ \ \ \ \ \ m\ge 1
\ee
and choosing ${\cal N}=2N- \aleph$, one arrives at the partition function $Z^+_{BGW}$
that does not depend on $N$ (but only on the parameter $\aleph$)
though the integrand in (\ref{UBGW}) does!

\subsubsection{Determinant representation and integrability}

One can easily integrate over the angular variables in (\ref{UBGW}) as in the previous section
to obtain
\be
Z_{UBGW}={\det F_i(M_j)\over \Delta (M)}
\ee
where
\be
F_i(M)=\int dhh^{i-1}\exp\Big(-{1\over h}+MH-{\cal N}\log h\Big)=
2\pi i \left(2\sqrt{M}\right)^{{\cal N}-i}I_{{\cal N}-i}\left(2\sqrt{M}\right)
\ee
where $I_k(z)$ are the modified Bessel functions. After some work \cite{GKMU},
this formula can be recast to the form
(\ref{detGKM}) with the asymptotics (\ref{asym}), where $L=1/M$ and
\be
\phi_i(M)={\Gamma \left(2N-{\cal N}-2i+2\right)\over 2^{2-i}\pi i}
\left({\sqrt{\mu}\over 2}\right)^{2N-{\cal N}-1}I_{N-{\cal N}/2-i}\left({2\over\sqrt{\mu}}
\right)
\ee
Again in order to make these functions independent of $N$, one has to choose
${\cal N}=2N-\aleph$. At the same time, this proves that, under such a choice,
$Z^+_{UBGW}$ is a $\tau$-function of the KP hierarchy.

\subsection{Kontsevich phase}

In the Kontsevich phase the unitary matrix integral is considered as a function of variables
\be\label{timesK}
\tau_k=-{1\over k}\tr M^{-k}
\ee
This time the integral (\ref{PKBGW}) does not depend on $N$ provided it is
properly normalized:
\be\label{BGWK}
Z^+_{BGW}=e^{-2\ \tr M^{1/2}}\sqrt{{
\det
\left(M^{1/2}\otimes M +
M \otimes M^{1/2}\right)\over \left(\det M\right)^{N}}}Z_{BGW}
\ee
One can check by a direct (quite involved) calculation \cite{GKMU} that $Z^+_{BGW}$
depends only on odd times $\tau_{2k+1}$.

\subsubsection{Virasoro constraints}

Using the Ward identity (\ref{WI-BGWM-M})
one can now make the change to variables (\ref{timesK}) to
obtain the Virasoro constraints satisfied by $Z^+_{BGW}$ \cite{GN,GKMU}:
\be\label{VircoU}
\hat {\cal L}_m Z^+_{BGW} = 0,\ \ \ \ \ \ m\ge 0\\
\hat {\cal L}_m=-{1\over 2}\sum_{odd\ k}k\tau_k{\p\over\p \tau_{k-2m}}+{1\over 2}\sum_{{odd\ a,b}\atop{a+b=2m}}
{\p^2\over\p \tau_a\p \tau_b}+{\p\over\p \tau_{2m+1}}+{\delta_{m,0}\over 16}
\ee

\subsubsection{Determinant representation and integrability}

In the Kontsevich phase, performing integration over the angular variables
in (\ref{PKBGW}) and taking into account the normalization factor, one
obtains \cite{GKMU} the determinant representation (\ref{detGKM}) with the
asymptotics (\ref{asym}), where
\be
\phi_{i-N}(M)=2\sqrt{\pi}e^{-2M}M^{i-N-1/2}I_{i-N-1}(2M)
\ee
This means that the partition function $Z^+_{BGW}$ of the unitary matrix model
is a $\tau$-function of the KP hierarchy. Moreover, as it was already noted, it does not
depend on odd times and is, in fact, a $\tau$-function of the KdV hierarchy.

\subsection{Genus expansion and the first multiresolvents}

\paragraph{Generic phase.}
As before, we shift first time variables $\tau_{2k+1}\to\tau_{2k+1}+T_k$,
\be
\hat\nabla(z)=\sum_{k=0}^\infty \frac{1}{z^{k+1}}\frac{\p}{\p \tau_k}
\ee
\be
W'(z)=\sum_{k=0}^{n+1}\left(k+\frac{1}{2}\right)T_k z^k
\ee
\be
v'(z)=\sum_{k=0}^{\infty}\left(k+\frac{1}{2}\right)\tau_k z^k
\ee
and obtain the loop equations
\be
W'(z)\rho(z)=\rho^2(z)+f_{BGW}(z)+g^2\hat\nabla(z)\rho(z)+P_z^-\left[ v'(z)\rho(z)\right]+\frac{g^2}{16z}
\ee
\be
f_K(z)=P^+_z\left[W'(z)\rho(z)\right]=g^2 \hat R_K(z)\log Z
\ee
\be
\hat R_K(z)=-\sum_{k=1}^{n+1}\sum_{m=0}^{k-1}\left(k+\frac{1}{2}\right)T_kz^{k-m-1}\frac{\p}{\p T_m}
\ee
\be
\rho(z)=g^2\hat\nabla(z)\log Z
\ee

\be
W'(z)\rho_W^{(p|m+1)}(z,z_1,\ldots,z_m) -
f_W^{(p|m+1)}(z|z_1,\ldots,z_m) = \nn \\ =
\sum_q \sum_{m_1+m_2=m}
\rho_W^{(q|m_1+1)}(z,z_{i_1},\ldots,z_{i_{m_1}})
\rho_W^{(p-q|m_2+1)}(z,z_{j_1},\ldots,z_{j_{m_2}}) + \nn \\ +
\sum_{i=1}^m \left(z_i\frac{\partial}{\partial z_i}+\frac{1}{2}\right)
\frac{\rho_W^{(p|m)}(z,z_1,\ldots,\check z_i,\ldots,z_m) - \rho_W^{(p|m)}(z_1,\ldots,z_m)}{z-z_i} +
\hat\nabla(z)\rho_W^{(p-1|m+1)}(z,z_1,\ldots,z_m)+\nn\\
\frac{\delta_{p,1}\delta_{m,0}}{16z}
\ee

\paragraph{Gaussian phase.} In this case,
 $W'(z)=1$, $f^{(k|m)}=0$
\be
\rho^{(0|1)}(z)=0
\ee
\be
\rho^{(0|2)}(z_1,z_2)=0
\ee
\be
\rho^{(0|3)}(z_1,z_2,z_3)=0
\ee
\be
\rho^{(1|1)}(z)=\frac{1}{16z}
\ee
\be
\rho^{(1|2)}(z_1,z_2)=\left(z_2\frac{\p}{\p z_2}+\frac{1}{2}\right)
\frac{\rho^{(1|1)}(z_1)-\rho^{(1|1)}(z_2)}{z_1-z_2}
\ee
\be
\rho^{(2|1)}(z)=\left(\rho^{(1|1)}(z)\right)^2+\rho^{(1|2)}(z,z)
\ee
\be
\rho^{(1|2)}(z_1,z_2)=\frac{1}{32z_1z_2}
\ee
\be
\rho^{(2|1)}(z)=\frac{9}{256z^2}
\ee

\paragraph{Free energy in the Gaussian case.}
As a direct corollary of Virasoro constraints (\ref{VircoU}), one can calculate the free
energy expansion in the parameter $g$, $\log Z_{BGW}=\sum_{k=0} g^{2k-2}{\cal F}^{(k)}_{BGW}$:
\be
{\cal F}^{(0)}_{BGW}=0\\
{\cal {F}}^{(1)}_{BGW}=-1/8\,\ln  \left( \tau_{{0}}-2 \right)\\
{\cal {F}}^{(2)}_{BGW}=-{\frac {9}{32}}\,{\frac {\tau_{{1}}}{
(\tau_{{0}}-2)^{3}}}\\
{\cal {F}}^{(3)}_{BGW}= -{\frac {225}{64}}\,{\frac {\tau_{{2}}}{(\tau
_{{0}}-2)^{5}}}+{\frac {567}{64}}\,{\frac {{\tau_{{1}}}^{2}}{(\tau_{{0}}-2)^{6}}
} \\
{\cal {F}}^{(4)}_{BGW}=  -{\frac {55125}{512}}\,{\frac {\tau_{{3}}}{(\tau_{
{0}}-2)^{7}}}+{\frac {388125}{512}}\,{\frac {\tau_{{2}}\tau_{{1}}}{(\tau_{{0}}-2)^{
8}}}-{\frac {64989}{64}}\,{\frac {{\tau_{{1}}}^{3}}{(\tau_{0}-2)^{9}}}
\label{freeBGW}
\ee
In general ${\cal F}^{(p)}_{BGW}$ is a {\it polynomial}
\be
{\cal F}^{(p)}_{BGW} =
\sum_{k_1+\ldots +k_m=p-1}
c_{k_1,\ldots,k_m} Q_{k_1}\ldots Q_{k_m}
\ee
of the variables $Q_k = \frac{\tau_k}{(\tau_0-2)^{2k+1}}$.
This is the best illustration of drastic simplicity of
the BGW partition function as compared to the Kontsevich and
Hermitian cases, where all ${\cal F}^{(p)}$ are
sophisticated transcendental functions, and are simplified only in terms of
{\it moment} variables. One may say in the BGW case the moment variables are extremely simple.

\pagebreak

\part{Decomposition formulas \label{decof}}

\setcounter{section}{0}
\setcounter{equation}{0}
\def\theequation{\Roman{part}.\arabic{equation}}

\section{The idea of decomposition formulas \cite{AMMIM}}

The key observation is that multiresolvents
-- if defined according to the rule  (\ref{muddef}) --
are polydifferentials on the bare spectral curve $\Sigma$,
intimately related to the $\widehat{U(1)}$ current $\hat{\cal J}(z)$
on $\Sigma$, with prescribed singularities: usually they are
allowed at some fixed points (\textit{punctures}) on $\Sigma$.
In this approach the Virasoro constraints on partition function
are written as
\be
\hat{\cal P}_-\left(\hat{\cal J}^2(z)\right) Z =
\oint_C {\cal K}(z,z') \left(\hat{\cal J}^2(z)\right) Z = 0
\ee
with a certain kernel ${\cal K}(z,z')$, made out of the
free-field Green function on $\Sigma$.
The current is also "shifted":
$\hat{\cal J}(z) \longrightarrow \hat{\cal J}(z) + \Delta\hat{\cal J}(z)$
and partition function $Z$ depends on the choice of:

$\bullet$ the complex curve (Riemann surface) $\Sigma$,

$\bullet$ the Green function ${\cal K}(z,z')$, i.e. projection operator
$\hat{\cal P}_-$,

$\bullet$ the punctures on $\Sigma$ and associated loop operator
$\hat{\cal J}(z)$,

$\bullet$ the local coordinates in the vicinity of the punctures,

$\bullet$ the involution of the curve with punctures and loop operator,

$\bullet$ the shift $\Delta{\cal J}(z)$ on $\Sigma$,

$\bullet$ the contour $C$ which separates two sets of punctures.

If contour $C$ goes around an isolated puncture, $Z$ is actually
defined by its infinitesimal vicinity and depends on behavior
(the type of singularity) of $\hat {\cal J}(z)$ at this particular
puncture. Coordinate dependence is reduced to the action of
a unitary operator (Bogoliubov transform, and exponential of bilinear
function of $\hat{\cal J}$) on $Z$.
Types of singularities and associated $Z$'s can be classified,
and our quartet $Z_H$, $Z_C$, $Z_K$ and $Z_{BGW}$ are the lowest
members of this classification. The former two are associated with
a puncture at regular point of $\Sigma$, while the latter two --
with that at a second-order ramification point.
$Z_C$ and $Z_{BGW}$ differ from $Z_H$ and $Z_K$ by the choice
of projection operator $\hat{\cal P}_-$, i.e. the kernel ${\cal K}(z,z')$.

If contour $C$ is moved away from the vicinity of the puncture,
it can be decomposed into contours encircling all other punctures:
this provides relations between $Z$'s of different types,
associated with different punctures.
If $\Sigma$ has handles or boundaries, there will be additional
contributions, associated with non-contractible contours --
the corresponding elementary partition functions are not yet
identified and investigated -- this seems to be a very interesting
problem of its own.

In what follows we present the two simplest examples of this
procedure, both associated with $\Sigma$, represented as a double-covering
of the Riemann sphere with two ramification points.
Such $\Sigma$ is of course also a Riemann sphere, however, representing
it as a double-covering provides a simple description of behavior,
which we allow $\hat{\cal J}(z)$ to have at the two ramification points.
The other pair of punctures are chosen at preimages of a regular points
($z=\infty_\pm$ in what follows).
After that, depending on the choice of projection operator
${\cal P}_-$ we obtain either a relation between $Z_H$ and the two
Kontsevich models, $Z_H = \hat U_{KK} \Big(Z_K\otimes Z_K\Big)$,
or between $Z_C$ and the pair: Kontsevich model and BGW model,
$Z_C = \hat U_{KBGW} \Big(Z_K\otimes Z_{BGW}\Big)$.
These both examples were already described in \cite{AMMIM}, but here
we provide a more targeted and, hopefully, more clear presentation
of the subject. Some mistakes of original version are also corrected,
in the case of discrepancies from \cite{AMMIM} the present version
should be trusted more.

\section{The basic currents, shifts and projection operators
\label{basicc}}

These are the data, defining the standard Virasoro constraints
(\ref{dico}) and (\ref{coco}) and thus the four models, discussed in the
section \ref{4mamo} above. All the four are defined in vicinity of
a particular puncture and do not depend on the global properties
of the bare spectral curve $\Sigma$.

\subsection{Hermitian current}

This one is used in the definition of (\ref{divi})
and thus of partition functions $Z_H$ and $Z_C$.
\be
\hat J_H(z|g^2) = d\hat \Omega_H(z)=\sum_{k=0}^\infty \left (\frac{k}{2} t_k z^{k-1}dz
+ g^2 \frac{dz}{z^{k+1}}\frac{\partial}{\partial t_k}\right)
\label{hcur}
\ee
With this current one can immediately associate a
bi-differential $f_J(z,z')=\hat J(z)\hat J(z')-:\hat J(z)\hat J(z'):$
where the normal ordering means
all $t_k$ placed to the left of all $t$-derivatives.
It is related to the central extension $\widehat{U(1)}$
and is equal to
\be
f_H(z,z'|g^2)=g^2\frac{dz dz'}{2(z-z')^2}
\ee
This bi-differential will play an important role
in comparison of global and local currents and,
therefore, in construction of conjugation operators in the next subsections
\ref{H} and \ref{CKBGW}.

The further difference between various partition functions comes from
different choices of the shift functions $W(z)$ \cite{spef}
and projector operators \cite{AMMIM},
\be
P_{m}\left[\sum_{k=-\infty}^\infty \frac{a_k dz^2}{z^{k+2}}\right]
=\sum_{k=m}^\infty \frac{a_k dz^2}{z^{k+2}}
\ee
The two correlated choices lead to the two simplest models,
associated with (\ref{hcur}): to $Z_H$ an $Z_C$.

\begin{enumerate}
\item{\textbf{Gaussian Hermitian model \cite{spef}}}

This model corresponds to the shift
\be
\Delta \hat J_H(z)=-\frac{z dz}{2}
\label{hshi}
\ee
Partition function is completely fixed by Virasoro constraints
\be
\hat T_H(z) Z_{H}=0
\label{HVir}
\ee
where
\be
\hat T_H(z) = P_{-1}\left[:(\hat J_H(z)+\Delta \hat J_H(z))^2:\right]=
g^2\sum_{n=-1}\frac{(dz)^2}{z^{n+2}}\hat L_n, \nn\\
\hat L_n = \sum_{k=1}^\infty k\left(t_k-\frac{\delta_{k,2}}{2}\right)\frac{\partial}{\partial t_{k+n}}
+ g^2\sum_{k=0}^n\frac{\partial^2}{\partial t_k \partial t_{n-k}}
\ee
with
\be
\frac{\p}{\p t_0}Z_H(t)=\frac{S}{g^2}Z_H(t)
\ee
Given (\ref{hshi}), the choice of $P_{-1}$ from all the $P_m$ is distinguished:
with this choice only partition function is \textit{unambiguously}
defined by (\ref{HVir}).
There are interesting situations, when the choice of $P_m$ is not adjusted
to the shift in this way: the best known example is provided by
Dijkgraaf-Vafa partition functions \cite{DV,DVfol}, where projector is the same
$P_{-1}$ as in Gaussian model, but the shift $\Delta \hat J_H(z)=dW(z)$ is generated by polynomial
$W(z)$ of degree higher than two.

\item\textbf{Gaussian complex model \cite{comamo}}

This model corresponds to the shift
\be
\Delta \hat J_C(z)=-\frac{dz}{2}
\ee
Thus partition function is completely fixed by Virasoro constraints
\be
\hat T_C(z) Z_{C}=0
\label{CVir}
\ee
where projector is taken to be $P_0$ -- again, to guarantee the
uniqueness of the solution to (\ref{CVir}), -- and
\be
\hat T_C(z) = P_{0}\left[:(\hat J_H(z)+\Delta \hat J_C(z))^2:\right]=
g^2\sum_{n=0}\frac{(dz)^2}{z^{n+2}}\hat L_n, \nn\\
\hat L_n = \sum_{k=1}^\infty k\left(t_k-\delta_{k,1}\right)\frac{\partial}{\partial t_{k+n}}
+ g^2\sum_{k=0}^n\frac{\partial^2}{\partial t_k \partial t_{n-k}}
\label{virc}
\ee
with
\be
\frac{\p}{\p t_0}Z_C(t)=\frac{S}{g^2}Z_C(t)
\ee
\end{enumerate}

\subsection{Kontsevich current}

This one is used in the definition of (\ref{covi})
and thus of partition functions $Z_K$ and $Z_{BGW}$,
\be
\hat J_K(\xi|g^2) = d\hat \Omega_K(\xi)= \sum_{k=0}^\infty\frac{1}{2}\left(k+\frac{1}{2}\right)\tau_k \xi^{2k}d\xi +
g^2\frac{d\xi}{\xi^{2k+2}}\frac{\partial}{\partial\tau_k}
\ee
is -- up to traditional but unimportant change of time-variables --
the even part of the current (\ref{hcur}).
However, associated central term bi-differential looks more sophisticated
(being the symmetric part of the bi-differential):
\be
f_K(\xi,\xi'|g^2)=g^2\frac{(\xi^2+\xi'^2)d\xi d\xi'}{4(\xi^2-\xi'^2)^2}
\ee

The simplest partition functions, associated with this current,
are Kontsevich $\tau$-function and BGW model.

\begin{enumerate}
\item \textbf{Kontsevich $\tau$-function \cite{Ko,GKM}}

This time the shift is
\be
\Delta \hat J_K=-\frac{\xi^2 d\xi}{2}
\ee
and the relevant projector is $P_{-2}$:
\be
\hat T_K(\xi) = P_{-2}\left[:\left(\hat J_K+\Delta \hat J_K\right)^2:\right]=
g^2\sum_{n=-1}^\infty\frac{(d\xi)^2}{\xi^{2n+2}}\hat {\cal L}_n, \nn\\
\hat {\cal L}_n = \sum_{k=0}^\infty\left(k+\frac{1}{2}\right)\left(\tau_k-\frac{2\delta_{k,1}}{3}\right)\frac{\partial}{\partial \tau_{k+n}}
+ g^2\sum_{k=0}^{n-1}\frac{\partial^2}{\partial \tau_k \partial \tau_{n-1-k}}+\frac{\delta_{n,0}}{16}+\frac{\delta_{n,-1}\tau_0^2}{16g^2}
\ee
Then $Z_K$ is uniquely defined by
\be
\hat T_K(\xi)Z_K=0
\ee

\item \textbf{Brezin--Gross--Witten model \cite{GKMU}}

Now the shift is
\be
\Delta J_{BGW}=-\frac{c d\xi}{4}
\ee
and
\be
\hat T_{BGW}(\xi) = P_{0}\left[:\left(J_K+\Delta J_{BGW}\right)^2:\right]=g^2\sum_{n=0}^\infty\frac{(d\xi)^2}{\xi^{2n+2}}{\cal L}_n, \nn\\
{\cal L}_n = \sum_{k=0}^\infty\left(k+\frac{1}{2}\right)\left(\tau_k-2\delta_{k,0}\right)\frac{\partial}{\partial \tau_{k+n}}
+ g^2\sum_{k=0}^{n-1}\frac{\partial^2}{\partial \tau_k \partial \tau_{n-1-k}}+\frac{\delta_{n,0}}{16}
\ee
with projector $P_0$ unambiguously specify $Z_{BGW}$ by
\be
\hat T_{BGW}(\xi)Z_{BGW}=0
\ee

\end{enumerate}

\section{Decomposition relation $Z_H \longrightarrow Z_K\otimes
Z_K$ \label{H}}

Now we can select a bare spectral curve $\Sigma$:
\be
y^2=z^2-a^2
\label{Hbsc}
\ee
select the punctures: at $z=\pm a$ and $z=\infty_\pm$,
and select the global current by allowing specific singularities at punctures:
\be
\hat {\cal J}(z|g^2) =
\sum_{k=0}\left(k+\frac{1}{2}\right)(A_k+zB_k)y^{2k-1}dz + g^2(C_k+zD_k)\frac{dz}{y^{2k+3}}
\label{globcurH}
\ee
Bi-differential for this current
\be
f_{\cal J}=g^2\frac{(z z'-a^2)dz dz'}{2(z-z')^2y(z)y(z')}
\ee
is defined by commutation relations
\be
C_k=a^2\frac{\p}{\p A_k}+\frac{k+1}{k+\frac{3}{2}}\frac{\p}{\p A_{k+1}},
\ \ \ \ D_k=\frac{\p}{\p B_k},
\ee
At punctures it is
equivalent  to the bi-differentials of the basic currents:
\be
f_{\cal J}(z,z') \stackrel{z\rightarrow \infty^{\pm}}{\sim} f_H (z,z') \nn\\
f_{\cal J}(z,z') \stackrel{z\rightarrow \pm a}{\sim} 4f_K (\xi_{\pm },\xi_{\pm }')
\ee
where $\xi_{\pm}$ are some local coordinates in the vicinity of ramification points $a$ and $-a$,
defined  respectively by
\be
z=a+\sum_{k=1}^\infty{\alpha^+_k\xi_+^{2k}}
\label{lochp}
\ee
and
\be
z=-a+\sum_{k=1}^\infty{\alpha^-_k\xi_-^{2k}}
\label{lochm}
\ee
The current (\ref{globcurH}) itself is equivalent to the currents from s.\ref{basicc}:
\be
\hat {\cal J}(z) \stackrel{z\rightarrow \infty^{\pm}}{\sim} \hat J_H (z) \nn\\
\hat {\cal J}(z) \stackrel{z\rightarrow \pm a}{\sim} 2\hat J_K (\xi)
\ee
Time-variables in parametrization of  the global current are related to local time as follows:
\be
t_k\sim\frac{2}{k}\oint\frac{\hat {\cal J}}{z^k}\nn\\
\frac{\p}{\p t_k}\sim\frac{1}{g^2}\oint z^k {\hat {\cal J}}
\ee
\be
\tau_k\sim\frac{2}{2k+1}\oint \frac{\hat {\cal J}}{\xi^{2k+1}}\nn\\
\frac{\p}{\p \tau_k}\sim \frac{1}{2g^2}\oint\xi^{2k+1}{\hat {\cal J}}
\ee
Global current is related to local currents by conjugation operators.
Conjugation operator at infinity is
\be
U_H=\frac{2}{g^4}\oint_\infty\oint_\infty\left(f_{\cal J}(z,z')-f_H(z,z')\right)
\hat\Omega_H(z)\hat\Omega_H(z')=\\
=\frac{1}{2g^2}\oint_\infty\oint_\infty \rho_{H}^{(0|2)}(z,z') v(z) v(z')
\ee
where $\rho_{H}$ is a bi-differential counterpart of the
two-point function of Gaussian Hermitian model
\be
\rho_{H}^{(0|2)}(z,z')=\frac{1}{g^2}\left(f_{\cal J}(z,z')-f_H(z,z')\right)=
\frac{1}{2(z_1-z_2)^2}\left(
\frac{z_1z_2-a^2}{y(z_1)y(z_2)} - 1\right)
\ee
At ramification points the conjugation operator is as follows
\be
\hat V_H=\frac{2}{g^4}\sum_{{i,j}=\pm}\oint_{a_i}\oint_{a_j}\left(f_{\cal J}(z,z')-
4\delta_{ij}f_K(\xi_i,\xi_j)\right)\hat\Omega_K(\xi_i)\hat\Omega_K(\xi_j)
\ee

To establish required Virasoro constraints one should shift the global current
\be
{\hat{\cal J}} \to {\hat{\cal J}}-\frac{y(z)dz}{2}
\ee
which leads to a shift of the conjugation operators:
\be
U_H \to U_H +\frac{2}{g^2}\oint \frac{z-y(z)}{2}\hat\Omega_H(z)dz=
U +\frac{1}{g^2}\oint \rho^{(0|1)}_{GH}(z)v(z)dz\\
\hat V_H \to \hat V_H +\frac{2}{g^2}\left(\oint_{\xi_+=0}\left(\frac{2\xi_+^2 d \xi_+
-y(z)dz }{2}\right)\hat\Omega_K(\xi_+)+\oint_{\xi_-=0}\left(\frac{2\xi_-^2 d\xi_-
-y(z)d z}{2}\right)\hat\Omega_K(\xi_-)\right)
\ee

Then the projector
\be
\oint_C \frac{1}{(z-z')dz'} :({\hat{\cal J}}(z')+\Delta {\hat{\cal J}}(z'))^2:
\ee
with contour $C$ encircles the segment ramification points $\pm a$ on the spectral curve
(but not the point $z$! so that always $|z|>|z'|$) do the job:
since
\be
\frac{1}{(z-z')dz'} = \sum_{k\geq 0}\frac{(z')^k}{z^{k+1}dz'}
\ee
it picks up the terms with $n\geq -1$ in infinity
and since
\be
\frac{1}{(z-z')dz'} = \sum_{k\geq 0}\frac{(\xi')^{2k-1}}{2\xi^{2k+2}d\xi'}
\ee
it picks up the terms with $n\geq -1$ for ramification points.

After all we get the decomposition formula\footnote{Actually,
as it was already indicated in \cite{AMMIM}, we get a whole family
of such formulas,
with infinite set of free parameters given by coefficients
$\alpha^\pm_k$ in (\ref{lochp}), (\ref{lochm}).}
\be
Z_H(t)=e^{U_H}e^{\hat V_H}Z_K(\tau_+)Z_K(\tau_-)
\ee

\section{Decomposition relation $Z_C \longrightarrow Z_K\otimes
Z_{BGW}$ \label{CKBGW}}

This decomposition formula
was the topic of s.8 of ref.\cite{AMMIM}, however,
it is described there in a too sketchy and partly
misleading form. Thus we provide here a more detailed
and careful presentation.\footnote{
In \cite{AMMIM} we considered decomposition formula for the complex model,
starting from the same spectral curve (\ref{Hbsc}) as for the
Gaussian Hermitian matrix model
\be
y_H^2=z^2-4S
\nn
\ee
with additional puncture in $z=0_\pm$.
However, the global current which we introduced was singular at $0_\pm$.
Actually, in notations of \cite{AMMIM}, the proper global current should be defined
on the curve
\be
y_c^2=z^2(z^2-4S)
\nn\label{quadcurv}
\ee
It is more natural to consider instead the current on
\be
y_C^2=\xi(\xi-4S)
\nn\label{lincurve}
\ee
of which the previous one is a double covering $z=\sqrt{\xi}$
as we do in the present text.
}

The bare spectral curve is
\be
y_C^2=z(z-4S)
\label{lin}
\ee
and the four punctures are chosen at $z=0,4S,\infty_\pm$.
Accordingly on this curve we define the global current
\be
{\cal J}(z)=\sum_{k=0}^\infty  \left(k+\frac{1}{2}\right)\left(A_k+zB_k\right)y_C^{2k-1}dz+
g^2\frac{dz}{y_C^{2k+3}}\left(C_k+zD_k\right)
\ee
with commutation relations
\be
C_k=8S^2\frac{\p}{\p A_k}-2S\frac{\p}{\p B_k}+\frac{k+1}{k+\frac{3}{2}}\frac{\p}{\p A_{k+1}}\nn\\
D_k=\frac{\p}{\p B_k}-2S\frac{\p}{\p A_k}
\ee
and the global bi-differential
\be
f_{\cal J}(z,z')=g^2\frac{(z z'-2S(z+z'))dzdz'}{2(z-z')^2y_Cy_C'}=\\
=g^2\frac{2y_C^2y_C'^2+(z z'-2S(z+z')+8S^2)(y_C^2+y_C'^2)}{2(y_C^2-y_C'^2)^2y_Cy_C'}dz dz'
\label{difffC}
\ee
At punctures this bi-differential
is equivalent to the following canonical bi-differentials from s.\ref{basicc}:
\be
f_{\cal J}(z,z')\stackrel{z\rightarrow \infty^{\pm}}{\sim} f_H(z,z')\nn\\
f_{\cal J}(z,z')\stackrel{z\rightarrow 4S}{\sim} 4f_K(z,z')\nn\\
f_{\cal J}(z,z')\stackrel{z\rightarrow 0}{\sim} 4f_K(z,z')
\ee
The current has the following behavior:
\be
{\cal J}(z) \stackrel{z\rightarrow \infty^{\pm}}{\sim} J_H (z) \nn\\
{\cal J}(z) \stackrel{z\rightarrow 4S}{\sim} 2J_K (\xi_+)\nn\\
{\cal J}(z) \stackrel{z\rightarrow 0}{\sim} 2J_K (\xi_-)\\
\ee
with $\xi_+$, $\xi_-$ -- some local coordinates in the vicinities of $4S$, $0$ respectively:
\be
z=4S+\sum_{k=1}^\infty{\alpha^+_k\xi_+^{2k}} \nn\\
z=\sum_{k=1}^\infty{\alpha^-_k\xi_-^{2k}}
\ee
Time-variables of the local currents are expressed through those of the global one
in the same way as in s.\ref{H}:
\be
t_k\sim\frac{2}{k}\oint\frac{\cal J}{z^k}\nn\\
\frac{\p}{\p t_k}\sim\frac{1}{g^2}\oint z^k {\cal J}
\ee
\be
\tau_k\sim\frac{2}{2k+1}\oint \frac{\cal J}{\xi^{2k+1}}\nn\\
\frac{\p}{\p \tau_k}\sim \frac{1}{2g^2}\oint\xi^{2k+1}{\cal J}
\ee
Global current is related to local currents through conjugation operators.
Conjugation operator at infinity
\be
U_C=\frac{2}{g^4}\oint_\infty\oint_\infty\left(f_{\cal J}(z,z')-f_H(z,z')\right)\Omega_H(z)\Omega_H(z')=\\
=\frac{1}{2g^2}\oint_\infty\oint_\infty \rho_{C}^{(0|2)}(z,z') v(z) v(z')
\ee
where $\rho_{C}$ is a bi-differential counterpart of
the two-point function of Gaussian Hermitian model
\be
\rho_{C}^{(0|2)}(z,z')=\frac{1}{g^2}\left(f_{\cal J}(z,z')-f_H(z,z')\right)
\ee
At ramification points the conjugation operator looks as follows
\be
V_C=\frac{2}{g^4}\sum_{{i,j}=\pm}\oint_{a_i}\oint_{a_j}\left(f_{\cal J}(z,z')-4\delta_{ij}f_K(\xi_i,\xi_j)\right)\Omega_K(\xi_i)\Omega_K(\xi_j)
\ee

The shift of the current
\be
{\cal J} \to {\cal J} -\frac{y_C(z) dz}{2z}
\ee
corresponds to the shift of conjugation operators
\be
U_H \to U_H +\frac{2}{g^2}\oint \frac{z-y(z)}{2z}\Omega_H(z)dz=U +\frac{1}{g^2}\oint \rho_{C}^{(0|1)}(z)v(z)dz\\
V_H \to V_H +\frac{2}{g^2}\left(\oint_{\xi_+=0}\left(\xi_+^2 d \xi_+ -\frac{y(z)dz }{2z}\right)\Omega_K(\xi_+)+\oint_{\xi_-=0}\left(\frac{c d\xi_-}{2} -\frac{y(z)d z}{2z}\right)\Omega_K(\xi_-)\right)
\ee

The difference from the case of Hermitian model is that now we should get
\be
L_n,\ \ n\geq 0 \ \ {\rm at} \ \infty, \nn \\
{\cal L}_n, \ \ n\geq 0\ \ {\rm at} \ a, \nn \\
{\cal L}_n, \ \ n\geq -1\ \ {\rm at} \ -a
\ee
Thus this time the proper projector is
\be
\oint_C \frac{z'}{(z-z')dz'} :({\cal J}(z')+\Delta {\cal J}(z'))^2:
\ee
an we finally obtain the decomposition formula for complex model:
\be\label{dcfbgw}
Z_{C}(t)=e^{V_C}e^{U_C}Z_K(\tau_+)\tilde{Z}_K(\tau_-)=e^{V_C}
e^{U_C}Z_K(\tau_+){Z}_{BGW}(\tau_-)
\ee

\section*{Conclusion \label{concl}}

In this paper we demonstrated that decomposition formula
$Z_H \rightarrow Z_K\otimes Z_K$ of partition function for Gaussian
Hermitian model into two cubic Kontsevich models has as its closest
analogue another decomposition: $Z_C \rightarrow Z_K\otimes Z_{BGW}$
of the Gaussian complex model into the cubic Kontsevich and
Brezin-Gross-Witten models.
Thus all the four models are indeed the very close relatives,
though this is not quite so obvious from their original matrix-integral
representations.
This paper is therefore an important outcome and summary of many different
approaches, worked out during the years of development of matrix-model theory.
It brings us one-step closer to providing a unified look at the
whole variety of eigenvalue models and building up the
\textit{M-theory of matrix models}, suggested in \cite{AMMMT}.

Technically it adds to content of \cite{AMMIM} an identification
of partition function, denoted there by $\tilde Z_K$, with that of
the very important BGW model -- the generating function of all
unitary-matrix correlators.  From technical point of view the road
is now open for search of two different generalizations:
to Dijkgraaf-Vafa models \cite{DV,DVfol}, which are not \textit{fully} specified
by the Virasoro constraints alone and rely upon intriguing
and under-developed theory of \textit{check-operators} \cite{AMMcheck},
and to more interesting unitary-matrix models with Itzykson-Zuber measures
and further to Kazakov-Migdal multi-matrix models \cite{latthe}-\cite{ShaMo},
important both for Yang-Mills theory and for the theory of integer
partitions.  Putting all these very different problems into the same
context, moreover, underlined by the well established theory of
free fields on Riemann surfaces \cite{intappl}, is a challenging
and a promising perspective.

Another, but, perhaps, related,
open problem is {\it direct} derivation of decomposition
formula (\ref{dcfbgw}) from integral representations of
all the models, bypassing the Virasoro constraints and
$D$-module representations.
Note that this kind of problem remains unsolved even for the
crucially important decomposition
$Z_H = \hat U(Z_K\otimes Z_K)$,
describing the double-scaling continuum limit of Hermitian
matrix model.

\section*{Acknowledgements}

A.A. is grateful to Denjoe O'Connor for his kind hospitality while
this work was in progress. Our work is partly supported by Russian
Federal Nuclear Energy Agency, by the Dynasty Foundation (A.A.), by
the joint grants 09-02-91005-ANF, 09-02-90493-Ukr, 09-02-93105-CNRSL
and 09-01-92440-CE, by the Russian President's Grant of Support for
the Scientific Schools NSh-3035.2008.2, by RFBR grants 08-01-00667
(A.A.), 07-02-00878 (A.Mir.) and 07-02-00645 (A.Mor.).

\end{document}